\pdfoutput=1

\documentclass[11pt,a4paper]{revtex4}
\usepackage{epsfig}
\usepackage{graphicx}
\usepackage[titletoc]{appendix}
\usepackage{enumerate}
\usepackage{epsfig}
\usepackage{dcolumn}
\usepackage{bm}
\usepackage{amsmath}
\usepackage{amsfonts}
\usepackage{amssymb}
\usepackage{graphicx}
\usepackage{latexsym}
\usepackage{rotating}
\usepackage{multirow}
\usepackage{slashed}
\usepackage{xcolor}
\usepackage{hyperref}

\begin{document}
\title{NMSSM explanations of the Galactic center gamma ray excess and promising LHC searches}

\author{Jun Guo$^1$, Jinmian Li$^2$, Tianjun Li$^{1,3}$ and Anthony G.\ Williams$^2$}

\affiliation{
$^1$ State Key Laboratory of Theoretical Physics, Institute
of Theoretical Physics, Chinese Academy of Sciences, Beijing 100190,
P. R. China  \\
$^2$ ARC Centre of Excellence for Particle Physics at 
the Terascale, Department of Physics, University of Adelaide, 
Adelaide, SA 5005, Australia \\
$^3$ School of Physical Electronics, University of
Electronic Science and Technology of China, Chengdu 610054, P. R.
China
}

\begin{abstract}
The Galactic Center Excess (GCE) is explained in the framework of the Next-to-Minimal Supersymmetric Standard Model (NMSSM) with a
$Z_3$ discrete symmetry. 
We show that  a resonant CP-odd Higgs boson with mass twice that of the Dark Matter (DM) candidate is favoured. Meanwhile, the DM candidate
is required to have relatively large coupling with the $Z$ boson through its Higgsino component in order to obtain correct
DM relic density.
Its LHC discovery potential via four signatures is discussed in detail. We find that the most sensitive signals are 
provided  by the Higgsino-like chargino and neutralino pair production with their subsequent decays 
into $W$ bosons, $Z$ bosons, and DM. The majority of the relevant parameter space can be probed
at the Large Hadron Collider with a centre-of-mass energy of 14 TeV and an integrated luminosity 1000 fb$^{-1}$.
 \end{abstract}

\maketitle
\section{Introduction}

The only current empirical evidence for the existence of Dark Matter (DM) is from gravitational effects, 
for example, the rotation curves of spiral galaxies and studies of mass distributions through gravitational lensing.
However, it seems reasonable to expect that DM may have a common origin with Standard Model (SM) matter
since the ratio of the relic density of DM to the relic density of SM matter is approximately 5 to 1, i.e., a number of
order one. If the DM and SM origins were unrelated, then there would be no reason to expect this to occur.
If there is a common origin, perhaps in some Grand Unified Theory (GUT), then it seems reasonable to expect that 
there may be some non-gravitational interactions between DM and SM matter as well. There is however no 
clear and broadly accepted evidence for such interactions to date. In particular, the absence of any
convincing signals in spin independent DM direct detection 
experiements~\cite{Akerib:2013tjd, Agnese:2014aze} has already imposed very stringent upper limits 
on the couplings of DM particles to the SM particles.

DM can also be searched for through indirect signals arising from DM-DM annihilation into SM particles in regions of high
DM density. Such decays would potentially lead to anomalies in the energy spectra of gamma rays, anti-protons, electrons,
positrons and so on, where an anomaly would be something that could not be understood in terms of known astrophysical
sources. A recent analysis of the Fermi Gamma-Ray Space Telescope data~\cite{Daylan:2014rsa}, has shown a significant 
excess of gamma rays from the Galactic Center (GC). This GC  Excess (GCE) in the range of $\sim [1,3]$ GeV 
can be fitted very well by a 31-40 GeV DM particle annihilating into $b\bar{b}$ with an annihilation cross section 
of $\langle \sigma v \rangle = (1.4 \sim 2.0) \times 10^{-26}$ cm$^3$/s, or a 7-10 GeV DM particle annihilating into 
$\tau{\bar \tau}$ with similar cross section.
Potential explanations of the GCE have been studied extensively using both model-dependent and model-independent 
approaches~\cite{Alves:2014yha,Berlin:2014tja,Agrawal:2014una,Izaguirre:2014vva,Ipek:2014gua,Ko:2014gha,Abdullah:2014lla,Martin:2014sxa,Berlin:2014pya,Cline:2014dwa,Ghosh:2014pwa,Boehm:2014bia,Wang:2014elb,Agrawal:2014aoa,Han:2014nba,Huang:2014cla,Balazs:2014jla,Fields:2014pia,Cheung:2014lqa,Kong:2014haa,Bringmann:2014lpa,Cirelli:2014lwa,Okada:2014usa,Borah:2014ska,Cahill-Rowley:2014ora,Arina:2014yna,Hardy:2014dea,Banik:2014eda,Boehm:2014hva,Bell:2014xta,Modak:2013jya}.

It is well-known that supersymmetry (SUSY) provides a natural solution to the gauge hierarchy problem in the SM and, in
addition, SUSY leads to improved gauge coupling unification.
Furthermore, with the assumption of $R$-parity conservation the lightest supersymmetric particle (LSP)
such as the neutralino is stable and so is a good cold DM candidate. In short, SUSY
is one of the most promising candidates for new physics beyond the SM.
However, the Minimal SSM (MSSM) is challenged by the discovery of the 125 GeV Higgs 
boson~\cite{Aad:2012tfa,Chatrchyan:2012ufa} due to the so-called little hierarchy problem concerning the fine tuning 
at the electroweak scale.
In the Next-to-MSSM (NMSSM)~\cite{Ellwanger:2009dp}, the SM-like Higgs boson mass can be lifted by 
both tree-level coupling and the mixing with a lighter singlet~\cite{Ellwanger:2011mu, Ross:2011xv,Hall:2011aa,Kang:2012sy,Cao:2012fz,Cao:2012yn,Espinosa:2012in, Perelstein:2012qg, Agashe:2012zq,King:2012is,King:2012tr}. 
Thus, one can naturally obtain a relatively heavy Higgs boson. Moreover, the NMSSM has the advantage of also solving 
the $\mu$ problem in the MSSM, where the electroweak scale $\mu$ can naturally be generated by a singlet vacuum expectation value (VEV)
at $\mathcal{O}(100)$ GeV.  
In this paper, we demonstrate another merit of the NMSSM, which is the potential to explain the GCE elegantly through 
 a light singlet-like CP-odd Higgs.

Several studies of the GCE have been undertaken in the context of the NMSSM. DM pairs can annihilate into $b\bar{b}$ or $\tau \bar{\tau}$ via t-channel bottom squark or tau slepton exchange~\cite{Han:2014nba}. This process is, however, highly constrained by sparticles searches at LEP and LHC, by SM precision measurements as well as by DM direct detection searches. 
If the singlet-like scalar/pseudoscalar in the NMSSM is lighter than the DM particle, the DM pair annihilating into those new Higgs bosons that in turn have a dominant decay into bottom quarks ~\cite{Berlin:2014pya, Gherghetta:2015ysa}  will be able to give a promising fit to the gamma ray spectrum. This explanation requires an appropriate relationship between the final state Higgs boson mass and its coupling to DM and this can typically be realized within the general NMSSM framework. However, in the $Z_3$ NMSSM we find that the DM mass, the singlet scalar/pseudoscalar mass and their couplings are correlated and so there is less freedom to achieve this. We will briefly illustrate these difficulties concerning the $Z_3$ NMSSM later.  
A more promising channel is the $s$-channel pseudoscalar mediated annihilation~\cite{Cheung:2014lqa, Huang:2014cla,Cahill-Rowley:2014ora}. The annihilation cross section of this process can receive a Breit-Wigner enhancement when the pseudoscalar mass is approximately twice the DM mass where the resulting process is not suppressed by small DM velocity.  We will focus on this scenario in this work~\footnote{Several papers~\cite{Cao:2014efa,Gherghetta:2015ysa,Bi:2015qva} following the submission of our work have also used this mechanism to produce the GCE.}.
A right-hand sneutrino DM particle in the NMSSM can also provide a possible explanation of the GCE through its annihilation into bottom quark pair~\cite{Cerdeno:2014cda} or CP-even/CP-odd Higgs bosons pair~\cite{Cerdeno:2015ega}, while fulfilling other experimental constraints, however these studies are considerably more involved and we will not consider them further in this work.

The CERN Large Hadron Collider (LHC) is vigorously continuing its searches for evidence of SUSY. 
The current searches have the ability to exclude electroweakinos with masses below $\sim 700$ GeV~\cite{Aad:2014nua}. 
The next run of LHC at a centre-of-mass energy of 14 TeV will start in 2015. If SUSY is not found in this next run,
then it will provide much higher lower-mass bounds on all SUSY particles. Thus, whether or not we can probe the above 
NMSSM scenario at the 14 TeV LHC is an interesting question.
On the one hand, DM is expected to have relatively large annihilation cross section to
SM particles in order to reproduce both the correct DM relic density and the GCE. Hence, the
inverse process may help to produce significant numbers of DM pairs at the LHC if the DM mass is not too large.
On the other hand, the GCE scenarios in the NMSSM generically have light Higgsinos as well. So, probing the existence 
of a relatively light Higgsino can become a smoking gun for an explanation of the GCE in the NMSSM. We conclude 
that it will be difficult to search for the direct DM pair production with a mono-jet signature, whereas most of 
the viable parameter space is discoverable at 14 TeV LHC with 1000 fb$^{-1}$ of data by searching 
for $WZ$+DM final states decaying from Higgsinos.

This paper is organised as follows. In Section~\ref{DM}, based on the analysis in Ref.~\cite{Daylan:2014rsa}, 
we briefly introduce the gamma ray flux measurement at the galactic centre. We study in detail the accommodation of the GCE 
in the NMSSM in Section~\ref{NMSSM}. In Section~\ref{LHC}, we discuss some potential signatures 
and the discovery potential of the s-channel $A_1$ mediated annihilation scenario at the LHC. In particular, we investigate the required 
 data sample necessary to fully explore the viable parameter space. 
Finally, we summarize and present our conclusion in Section~\ref{conc}.

\section{Dark Matter Annihilation and GCE}
\label{DM}

Indirect detection, which focuses on DM annihilation into SM particle final states, is an important method to
search for non-gravitational evidence of DM.
The galactic centre is expected to have a relatively large DM number density due to gravitational effects
and so is the most promising place for dark matter indirect detection. 
Gamma ray signals of DM annihilation have advantages over the other indirect searches 
such as those involving anti-protons, positrons, and electrons, as gamma rays are neither deflected by
magnetic fields nor do they lose energy during their propagation.

The gamma-ray flux produced from DM annihilation near the galactic centre and then detected near the Earth 
is given by
\begin{align}
\frac{d \Phi}{d E_{\gamma}}=\frac{1}{4\pi}\frac{\langle\sigma v\rangle}{2m_{DM}^2}\frac{d N_{\gamma}}{d E_{\gamma}}\int_{\triangle \Omega}\langle J\rangle d\Omega^{'}~,~
\end{align}
where
\begin{align}
\langle J\rangle=\int_{\rm los} \rho^2(\textbf{r}) dl ~
\end{align}
is the so-called J-factor which
encapsulates the dark matter distribution integrated over a solid angle $\triangle \Omega$ along the
line-of-sight (los) and where
$d N_{\gamma}/d E_{\gamma}$ is the gamma-ray spectrum dependent on the properties  
of the final states of the DM annihilation and their kinematical features. Here, 
 $\rho(\bf r)$ is the assumed density distribution of the galactic DM halo. 

The Milky Way's DM density distribution is assumed to be approximately spherically symmetric and hence 
can be approximately described as a function $\rho(r)$ of distance $r$ from the Galactic Center. The NFW 
(Einasto and Navarro, Frenk and White) profile~\cite{Navarro:2008kc,Diemand:2008in} provides good fits 
to DM numerical simulations and is given by
\begin{align}
\rho(r)=\rho_0(\frac{r}{r_s})^{-\gamma}[1+(\frac{r}{r_s})]^{\gamma-3}~,~
\end{align}
where the scale parameter $r_s$, the density parameter $\rho_0$ and the 
central slope parameter $\gamma$ characterise the DM halo. The canonical
NFW value of the central slope parameter is $\gamma=1$. 
We follow Ref~\cite{Daylan:2014rsa} who use a scale parameter of $r_s=20$ kph
and choose $\rho_0$ so as to give
a dark matter density around the Sun 
of $\rho(r_{\rm sun})=0.3 ~\text{GeV cm}^{-3}$, where the distance of the Sun from the
Galactic Centre is $r_{\rm sun}=8.5$ kpc. This reference allows $\gamma$ to be a free parameter
and obtains a best fit value of $\gamma=1.26$.
For those DM indirect detections which focus on the Galactic Center  
the uncertainty of $\rho^2(r)$ can become extremely large very close to the center. 

The J-factor can be rewritten as follows
\begin{align}
J=\int db \int dl \int ds \cos b \rho(r)^2 ~,~
\end{align}
in which $r=(s^2+r_s^2-2s r_s\cos l\cos b)^{1/2}$ is the Galactocentric distance, and $(l,b)$ are
respectively the longitude and latitude angles. Ref.~\cite{Daylan:2014rsa} considered the gamma spectra 
within a $5^{\circ}\times5^{\circ}$ region around Galactic Center. 
$s$ is the line of sight distance which has to be integrated over.

The last and most important factor, the annihilation cross section 
$\langle\sigma v\rangle$, affects both indirect detection and DM relic density predictions, is 
a model dependent variable. The relic density is approximately given by
\begin{align}
\Omega h^2=\frac{m_{\chi} n_{\chi}}{\rho_{c}}\simeq \frac{3\times 10^{-27} cm^3 s^{-1}}{\langle\sigma v\rangle}~.~\,
\end{align}
At the freeze-out $v \sim 0.1$, one usually needs an annihilation cross section around 1 pb to produce 
the correct relic density $\Omega h^2\simeq 0.1$~\cite{Ade:2013zuv}.
The analysis of Ref~\cite{Daylan:2014rsa} showed that if the DM annihilates only into $b \bar{b}$ final states
 with $v \sim 10^{-3}$, one would need  $\langle\sigma v\rangle\simeq 2.2\times 10^{-26} ~\text{cm}^3 \text{s}^{-1}$ 
to produce the observed Gamma Ray Excess. Surprisingly, such annihilation cross sections can almost lead to  
the observed relic density. In the non-relativistic  limit,  the annihilation cross section can be written 
as $\langle\sigma v\rangle=a+ \mathcal{O}(b v^2)$. If a single process is used to explain the relic density and GCE simultaneously, the $a$
term should be non-vanishing and dominate the $\langle\sigma v\rangle$. 
\section{GCE in the NMSSM}
\label{NMSSM}

In this Section, we will apply the Galactic Center gamma-ray Excess to a phenomenological NMSSM 
with a $Z_3$ discrete symmetry. The corresponding superpotential is~\cite{Ellwanger:2009dp}
\begin{equation}
\label{sp}
W_{\text{NMSSM}}=h_u \hat{H}_u \hat{Q} \hat{U}^c_R + h_d \hat{H}_d \hat{Q} \hat{D}^c_R + h_e \hat{H}_d \hat{L} \hat{E}^c_R+ \lambda \hat{S} \hat{H}_u \hat{H}_d + \frac{\kappa}{3} \hat{S}^3
\,\,,\end{equation}
where $\hat{Q}$, $ \hat{U}^c_R$, $\hat{D}^c_R$, $\hat{L}$, $\hat{E}^c_R$, $\hat{H}_u$, $\hat{H}_d$ are
the superfields for quark doublet, right-handed up-type quark, right-handed down-type quark,
lepton doublet, right-handed charged lepton, up-type Higgs doublet, and down-type Higgs doublet
respectively.

Because of the absence of sparticles with mass equal to their SM partners, SUSY has to be broken 
at a high energy scale in the hidden
sector, and then the breaking effects are transmitted to the observable sector. The low energy 
supersymmetry breaking soft terms such as gaugino masses, scalar masses, and trilinear soft terms are
\begin{align}
-\mathcal{L}_{\text{soft}}= & {1\over 2} \left(M_3 \tilde{g} \tilde{g} + M_2 \tilde{W} \tilde{W}
+ M_1 \tilde{B} \tilde{B} \right) \cr
&+  m^2_{H_u}|H_u|^2  + m^2_{H_d}|H_d|^2 + m^2_S|S|^2 + m^2_Q|Q^2|+m^2_U|U^2_R| \cr
&  + m^2_D|D^2_R| + m^2_L|L^2| + m^2_E|E^2_R| \cr
&  + (h_uA_u Q H_u U^c_R -h_dA_dQH_dD^c_R - h_e A_e L H_d E^c_R \cr
&  + \lambda A_\lambda H_u H_d S + \frac{1}{3}\kappa A_\kappa S^3 + h.c.)\,\,,
\end{align}
where $\tilde{g}$, $\widetilde{W}$, and $\widetilde{B}$ are gluino, Wino, and Bino respectively.

The NMSSM has five neutralinos ($\tilde{\chi}^0_i$), which are the mass eigenstates of mixings among $\tilde{B}, \tilde{W}, \tilde{H}^0_d, \tilde{H}^0_u$, and $\tilde{S}$. Following the convention of Ref.~\cite{Ellwanger:2009dp}, we give the mass matrix for the neutralino sector in Appendix~\ref{NHmass}. Because of the extra singlet in the NMSSM, 
the neutral Higgs sector is also expanded. There are 3 CP-even Higgs ($H_i$) and 2 CP-odd Higgs ($A_i$), both of 
which are mixed among $H_d$, $H_u$ and $S$ gauge eigenstates.  We find it more convenient to discuss 
the CP-even and CP-odd Higgs mass matrixes in the Goldstone basis~\cite{Miller:2003ay},  
$S_i, (i=1,2,3)$ and $P_i, (i=1,2)$.  Their mass matrixes are presented in Appendix~\ref{NHmass} as well
 for later discussion.

Now, we are ready to discuss the feasibility of the scenarios that could fulfil the observation 
of the Galactic Center Gamma-ray Excess in the NMSSM, while remaining consistent with
 other experimental results. 

\subsection{$s$-Channel $A_1$ Resonant Annihilation}
\label{NMSSMs1}

As has been studied in Ref.~\cite{Daylan:2014rsa}, when DM annihilates directly into $b \bar{b}$, we need the DM 
to have mass $m_{\text{DM}} \sim 35$ GeV and  $\langle \sigma v \rangle |_{v\to 0} \sim 2 \times 10^{-26} ~\text{cm}^3/$s 
to produce the observed GCE.
Interestingly, the $b\bar{b}$ final state is commonly favoured due to its relatively larger Yukawa coupling
if the $s$-channel annihilation is mediated by a relatively light Higgs boson.
However, the Higgs boson mediator can not be CP-even primarily because of the following two reasons:
\begin{itemize}
\item Annihilation which is mediated by an $s$-channel CP-even Higgs is suppressed by small DM velocity. As discussed 
in Section~\ref{DM}, in this case, the dark matter will annihilate much faster at early times than it does today. 
So $\langle \sigma v \rangle |_{v\to 0} \sim 2 \times 10^{-26} ~\text{cm}^3/s$ will  lead to a very small relic density; and
\item The CP-even Higgs boson can also mediate spin-independent DM interactions with SM matter and
give a direct detection signal, which has a current bound as low as $10^{-9}$ pb~\cite{Akerib:2013tjd}.
This makes it very difficult to achieve a signal consistent with the GCE 
while satisfying the direct detection constraints.
\end{itemize}

In principle, the CP-odd Higgs boson can be a very good $s$-channel mediator candidate.
Firstly, there is no velocity suppression in its annihilation.
A fermion-antifermion pair with momentum~\cite{Kumar:2013iva}
\begin{align}
k_1^{\mu}&=(E_1,\overrightarrow{k})=(\sqrt{m_{DM}^2+\overrightarrow{k}^2},\overrightarrow{k}), \\
 k_2^{\mu}&=(E_2,-\overrightarrow{k})=(\sqrt{m_{DM}^2+\overrightarrow{k}^2},-\overrightarrow{k}),
 \end{align}
have the corresponding spinors
\begin{align}
u(k_1)&=\left(
         \begin{array}{cc}
           \frac{k_1\cdot\sigma+m_{DM}}{\sqrt{2(k_1^0+m_{DM})}}\zeta_1 & \frac{k_1\cdot\overline{\sigma}+m_{DM}}{\sqrt{2(k_1^0+m_{DM})}}\zeta_1 \\
         \end{array}
       \right)^T, \\
v(k_2)&=\left(
         \begin{array}{cc}
           \frac{k_2\cdot\sigma+m_{DM}}{\sqrt{2(k_2^0+m_{DM})}}\zeta_2 & \frac{k_2\cdot\overline{\sigma}+m_{DM}}{\sqrt{2(k_2^0+m_{DM})}}\zeta_2 \\
         \end{array}
       \right)^T.
\end{align}
The matrix element of dark matter annihilation with a pseudoscalar vertex $\overline{u}\gamma^5v A$ is
\begin{align}
\overline{u}\gamma^5v=-\frac{1}{\sqrt{(E_1+m_{DM})(E_2+m_{DM})}}[(E_1+m_{DM})(E_2+m_{DM})+\overrightarrow{k}^2](\zeta_1^{\dag}\zeta_2)~,~
\end{align}
where the first term is dominant and does not suffer from any velocity suppression.

Secondly, the interaction of DM and nucleon through $t$-channel CP-odd Higgs is spin dependent.
For the same vertex $\overline{u}\gamma^5u A$ the matrix element of the direct detection process is
\begin{align}
\overline{u}(p_1)\gamma^5u(p_2)&=\frac{1}{\sqrt{(p_1^0+m_{DM})(p_2^0+m_{DM})}}
\xi_1^{\dag}[(p_2^0+m_{DM})(\overrightarrow{p_1}\cdot\overrightarrow{\sigma})-(p_1^0+m_{DM})(\overrightarrow{p_2}\cdot\overrightarrow{\sigma})] \cr
&\sim 2(p_1-p_2)^i(\xi_1^{\dag}\hat{S}^i\xi_2)~,~
\end{align}
which only gives a spin-dependent amplitude. So, the stringent bound from spin-independent direct detection experimental
searches does not apply.

However, in a realistic NMSSM treatment, there are additional constraints that should be considered.
In order to have a relatively large DM annihilation cross section, the CP-odd Higgs boson can not be too heavy. 
Also, its couplings to the SM particles should be suppressed to evade the current collider searches.  Interestingly, 
a light singlet-like CP-odd Higgs ($A_1$) can meet the need. As a result, we will assume a light singlet superfield 
in our model. Also,  it is natural to require that our DM be singlino dominant~\footnote{Ref.~\cite{Cheung:2014lqa} showed
that DM can also be admixture of Bino and Higgsino. As we will discuss it briefly later, the 
Bino and Higgsino mixing scenario is similar to the singlino and Higgsino mixing scenario in many aspects.}.
 Its mass is approximately $m_{\tilde{\chi}^0_1} \sim 2 \kappa s$, which means
\begin{align}
\frac{\kappa}{\lambda} = \frac{m_{\tilde{\chi}^0_1}}{2 \mu_{\text{eff}}} \lesssim \frac{35}{200},
\label{s1dmmass}
\end{align}
where the second inequality is from the non-discovery of any chargino at the LEP experiment~\cite{lepchargino}. 
The requirement of having no Landau pole all of the way up to the GUT scale imposes the requirement that
\begin{align}
\sqrt{\lambda^2 + \kappa^2} \lesssim 0.5~,
\end{align}
which indicates that $\kappa$ can only be $\lesssim 0.1$.
The requirement of $A_1$ being singlet-like leads to a suppression of the $A_1 b \bar{b}$ coupling,
although it might be enhanced with the choice of a relatively large $\tan\beta$.
So both vertexes of $A_1 b \bar{b}$ and $A_1 \tilde{\chi}^0_1 \tilde{\chi}^0_1$ are suppressed and typically it is not possible to 
get the correct right relic density as well as a signal consistent with the GCE with such small couplings.
However, a resonant enhancement can resolve this problem: when the total invariant mass of the initial states
is close to the $A_1$ mass, 
the total annihilation cross section is enhanced by a Breit-Wigner factor
\begin{align}
R(s) =\frac{1}{(s-M_A^2)^2+\Gamma_A^2 M^2_A}\, .
\end{align}
The cross section of such process is given by
\begin{align}
\langle\sigma v\rangle=\frac{3}{8\pi}\sqrt{1-\frac{m_b^2}{m_{\tilde{\chi}^0_1}^2}}\frac{\kappa^2C_{A_1d}m_b^4\tan^2\beta}{(s-m_{A_1}^2)^2+\Gamma_{A_1}^2 m^2_{A_1}}~.~
\end{align}
Thus, we need to require the mass of the CP-odd Higgs $A_1$ to be approximately $70$ GeV to reproduce
the GCE signal.
A problem still exists if we want to accommodate the relic density simultaneously.
At the early stage of the Universe when DM is frozen out, the temperature of the Universe 
is around $m_{\tilde{\chi}^0_1}/20$~\cite{Jungman:1995df}.  So, the energy of the DM is slightly higher at freeze-out 
than today. The Breit-Wigner factor is very sensitive to the initial energy around the mediator mass. 
The $\langle \sigma v \rangle |_{v\to 0} \sim 2 \times 10^{-26} ~\text{cm}^3/$s today will lead to an over abundance of relic
DM because of the reduction of resonant enhancement.
To resolve this we can increase the annihilation cross section at freeze-out by including a small Higgsino component in the DM,
which will give an additional contribution from the $s$-channel $Z$ boson exchange. From Eqs.~(\ref{hdcomp}) and (\ref{hucomp}), 
we need small $\mu$ to get relatively large Higgsino components in DM. Moreover, the coupling between DM and the Z boson 
is proportional to tan$\beta$. So, a relatively large tan$\beta$ is required in our model as well. As will be shown 
numerically later, a $Z \tilde{\chi}^0_1 \tilde{\chi}^0_1$ coupling that can give the correct DM relic density can still
be consistent with the constraints on $Z$ boson invisible decays. The coupling 
between the DM and Higgs boson is enhanced by large $\lambda$ and, furthermore,
the $\tilde{H}_d$ component in DM is increased by large tan$\beta$. So $H_{\text{SM}}$ may have considerable decay
branching ratio to DM in this scenario.

Following the above arguments, we have found a viable DM scenario which can explain the relic density and GCE, and 
 is consistent with DM direct searches. 
Similar scenarios for singlino/Higgsino mixing DM have been studied in~\cite{Cheung:2014lqa}.  In their analysis, however, some experimental constraints, such as constraints on Higgs invisible decay,  are overlooked. We will present a concrete model below
which satisfies all of the current experimental constraints. Moreover, we have different conclusions regarding the
Bino/Higgsino mixing case and, as we will comment on later, the resonant DM annihilation enhancement can still be applied.  
 Before we present our the numerical scan,  let us study the parameter space more carefully 
to understand the correlations among the NMSSM  parameters.

The singlet tends to be lighter than the SM Higgs boson. Since the measurements of $H_{\text{SM}}$ are likely to be consistent
with the SM couplings, the mixing between the $H_u/H_d$ doublets and singlet should be suppressed in most cases. 
So we can arrive at an approximate value for $A_\lambda$, as shown in Eq.~(\ref{alambda}). Also, in the CP-even Higgs matrix, 
the $M^2_A$ is enhanced by a factor of $1/\text{sin}^2 2\beta$, which will increase greatly with large $\tan \beta$.  
So, we can safely decouple $S_1$ component in the CP-even Higgs matrix and the
$P_1$ component in the CP-odd Higgs matrix 
in the following discussions.  Then from the CP-odd Higgs boson mass matrix in Eq.~(\ref{pmass}) we obtain
\begin{align}
M^2_{A_1} \simeq -3 A_\kappa \frac{\kappa \mu}{\lambda} +   \frac{\lambda^2 v^2}{2 \mu} s_{2 \beta}(A_\lambda + \frac{4 \kappa \mu}{\lambda}) \simeq -3A_\kappa \frac{m_{\tilde{\chi}^0_1}}{2}+\lambda^2 v^2 (1+ \frac{\sin 2 \beta m_{\tilde{\chi}^0_1}}{2 \mu})~, \label{ma1}
\end{align}
where we have substituted Eq.~(\ref{alambda}) and $m_{\tilde{\chi}^0_1} \sim 2 \kappa s$.
Note that when $m_{\tilde{\chi}^0_1} \sim 35$ GeV, $\mu \gtrsim 100$ GeV, and tan$\beta$ is large,
the second term in the bracket of Eq.~(\ref{ma1}) can be ignored. As a result,  $A_\kappa$ is approximately fixed
to the value
\begin{align}
A_\kappa = \frac{2(\lambda^2 v^2 - M_{A_1}^2)}{3 m_{\tilde{\chi}^0_1}}~.~
\end{align}
Another possible problem is the Higgs invisible decay.
Although the direct bound on the SM-like Higgs invisible decay is still very weak, 
Br$(H_{\text{inv}}) \lesssim 75\%$~\cite{Aad:2014iia}, the current Higgs coupling measurement requires
 that the Higgs signal strength to the SM particles be around 1.
Because in most of our models, there is no additional Higgs production mechanism, the Higgs invisible decay 
branching ratio is thus highly constrained indirectly.
The main contribution to the vertex $H_2\tilde{\chi}^0_1\widetilde{\chi}^0_1$ is the superpotential
term $W=\lambda S H_u H_d$ and the corresponding coupling for  $H_2\tilde{\chi}^0_1\tilde{\chi}^0_1$ is
 proportional to $C_{\tilde{\chi}^0_1 d}\frac{\lambda}{\sqrt{2}}$, where $C_{\tilde{\chi}^0_1 d}$ is the $\tilde{H}_d$ component 
in the DM $\tilde{\chi}^0_1$. Hence we find
\begin{align}
\Gamma_{H_2\rightarrow\widetilde{\chi}^0_1 \widetilde{\chi}^0_1}=\frac{(1/4m_{H_2}^2-m_{\widetilde{\chi}^0_1})^{\frac{3}{2}} }{\pi m_{H_2}^2}\times\lambda^2 {C^2_{\widetilde{\chi}^0_1 d}}~.~
\end{align}
Thus, the Higgs invisible decay is enhanced by large $\lambda$ and $C_{\tilde{\chi}^0_1 d}$.

Let us summarize the results that we have so far:
\begin{itemize}
\item The $s$-channel mediator of DM annihilation should be a singlet-like CP-odd Higgs boson, whose mass is 
around $2\times m_{\text{DM}}$. This will fix the $A_\kappa$ parameter;
\item The singlino dominant DM should have a small Higgsino component, which requires a small $\mu$ and large $\tan \beta$.
Interestingly, natural SUSY requires a small $\mu$ as well;
\item The purity of the $H_{\text{SM}}$ can be fulfilled by setting an appropriate $A_\lambda$; and
\item The Higgs invisible decay limit needs to be carefully accounted for.
\end{itemize}

We use the NMSSMtools~\cite{Ellwanger:2004xm,Ellwanger:2005dv,Belanger:2005kh} software to survey the viable parameter space in
the NMSSM at the electroweak scale. The DM relic density and the DM direct and indirect detection rates are calculated 
using micrOMEGAs~\cite{Belanger:2013oya,Belanger:2010gh,Belanger:2008sj}. We apply the following constraints implicitly
 for our study:
\begin{itemize}
\item Theoretical constraints including converged RGE running, no tachyons, no Landau pole below the GUT scale, a physical 
global minimal and so on;
\item Limits from the Higgs and sparticle searches at the LEP and Tevatron experiments;
\item $B$ physics constraints;
\item Limits on the $Z$ boson invisible decay width;
\item The SM Higgs mass lies in the range of [123, 128] GeV and its signal strengths for all channels are lies in 
range of [0.8, 1.2] of the SM values; and
\item The requirement of a good dark matter candidate that has the properties
$0.09 < \Omega h^2 < 0.12$, $\sigma_{\text{SI}}< 1 \times 10^{-9}$ pb, and 
$0.5 \times 10^{-26} \text{cm}^3/s < \langle \sigma v \rangle < 5 \times 10^{-26}  \text{cm}^3/s$.
\end{itemize}
Within the scenario that we have proposed, to further simplify our scanning, we decouple the irrelevant sparticles by choosing
\begin{align}
M_1 = 1 ~\text{TeV},  M_2 = 2 ~\text{TeV},  M_3 = 3 ~\text{TeV}, m_{L} = m_E &= 1 ~\text{TeV}~, \nonumber \\
A_e = 0 ~\text{TeV}, m_{Q_2} =2 ~\text{TeV}, m_{Q_3} = 1.5 ~\text{TeV}, m_{U_2} = m_{D_2} &=3 ~\text{TeV}~, \nonumber \label{nmssmp1}\\
m_{U_3} = 1.5 ~\text{TeV}, m_{D_3} = 2.5~ \text{TeV}, A_b &= 0 ~\text{TeV} ~.~\,
\end{align}
We scan the remaining parameters in the following ranges
\begin{align}
\lambda: [0.35, 0.65], ~ \kappa: [30,45]\times \frac{\lambda}{2 \mu}, ~ \tan \beta: [10,40]~,~ \nonumber \\
\mu: [100,600] ~\text{GeV}, ~ A_{\kappa}: [-100,100] ~\text{GeV} ~,~\nonumber \\
A_{\lambda} = 2 \mu/\sin 2 \beta - 2 \kappa \mu /\lambda, ~ A_t = \mu/\tan \beta + \sqrt{6}\times1500 ~.~\,
\end{align}

\begin{figure}[htb]
  \centering
  \includegraphics[width=0.48\textwidth]{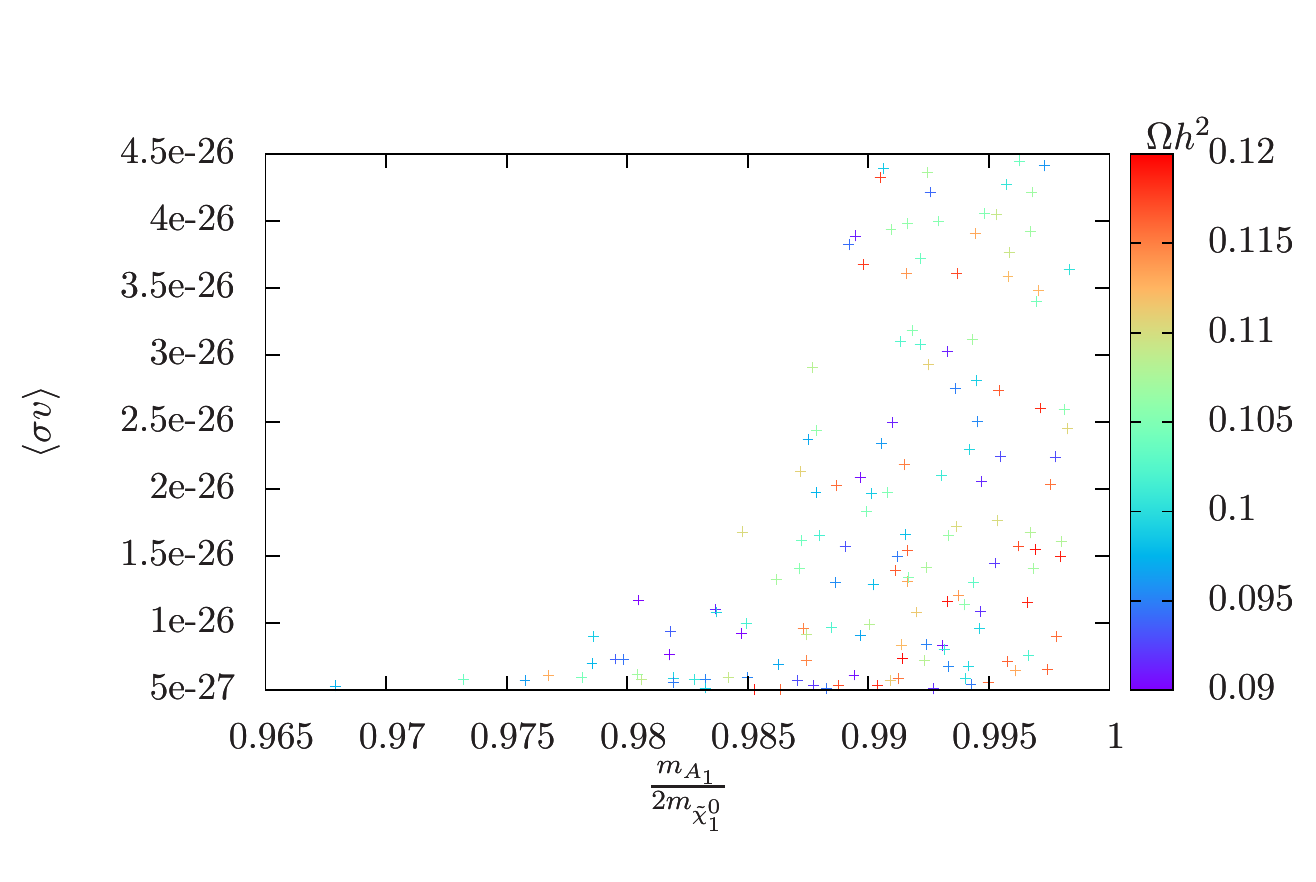}
  \includegraphics[width=0.48\textwidth]{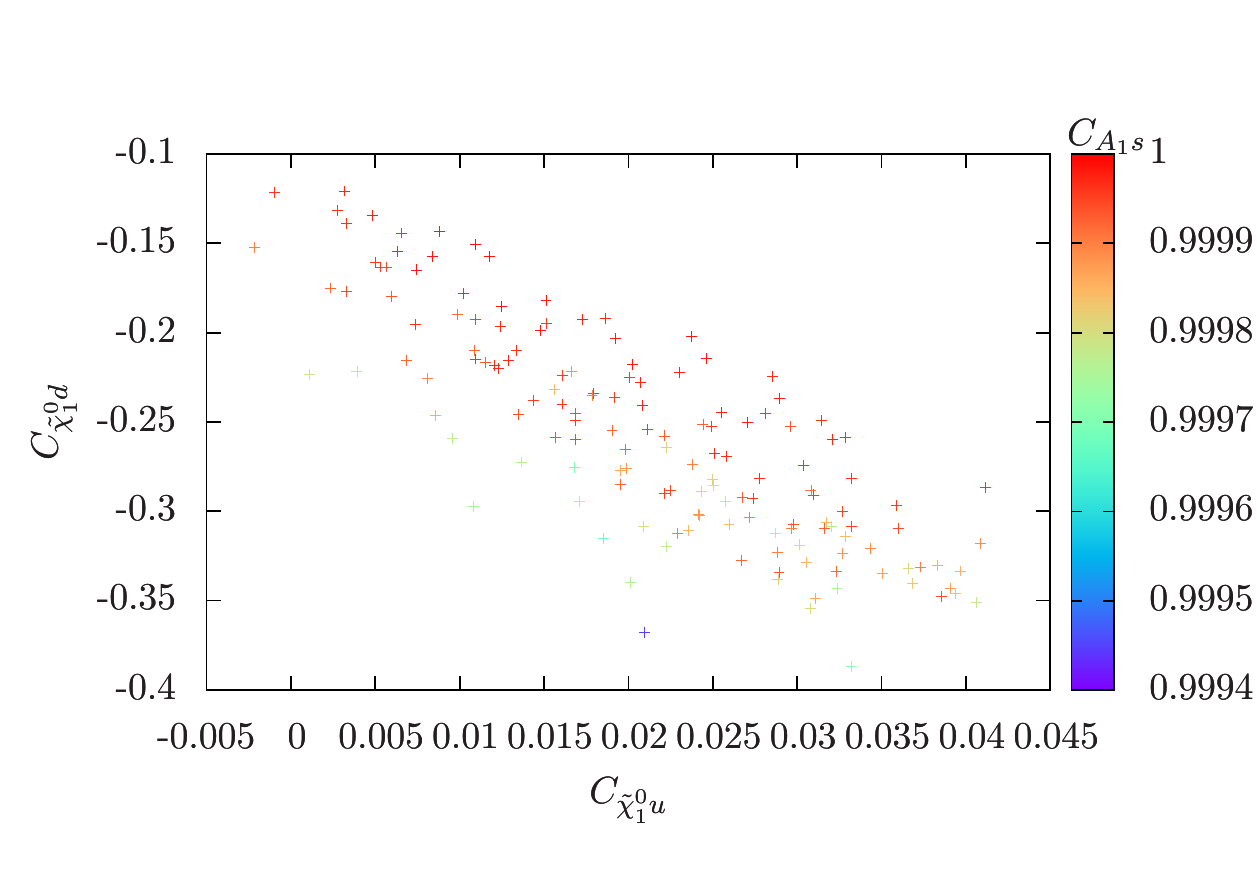} \\
  \includegraphics[width=0.48\textwidth]{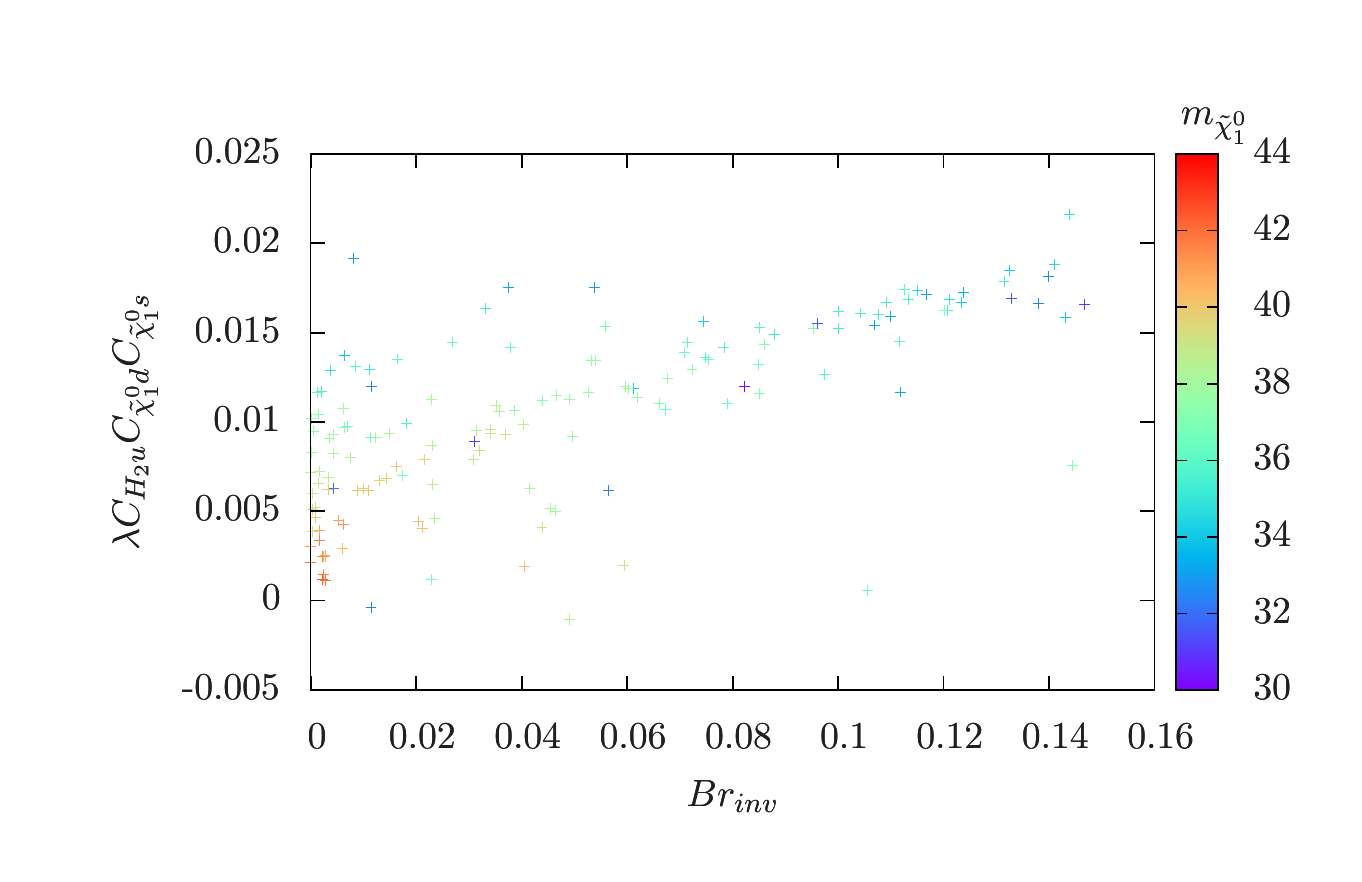}
    \includegraphics[width=0.48\textwidth]{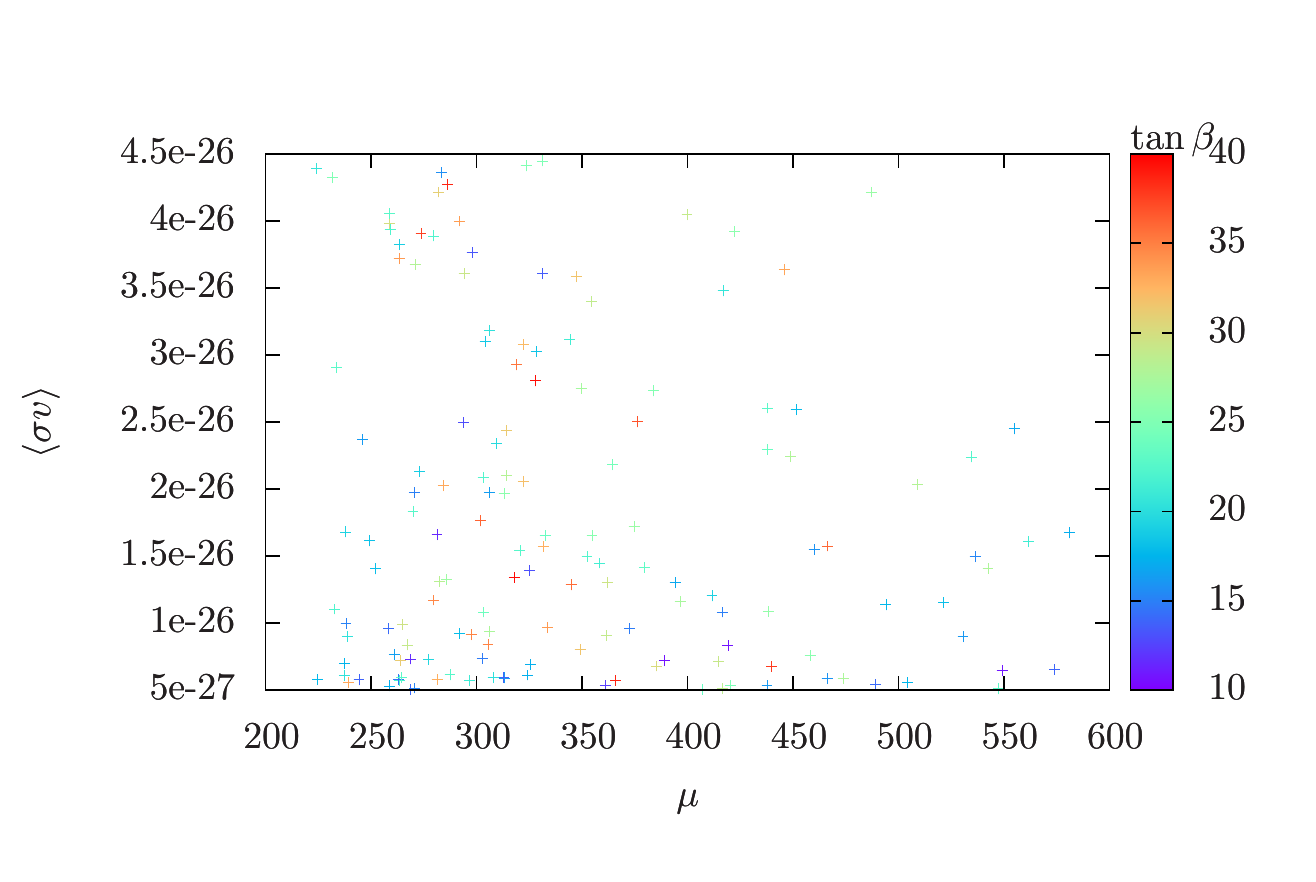}

  \caption{All the points in figures satisfy the constraints that mentioned in the Section~\ref{NMSSM}. 
The corresponding $x$-axis and $y$-axis as well as color coding are given in each panel. 
  \label{s1num}}
\end{figure}

Our numerical results are shown in Fig.~\ref{s1num}.
The DM annihilation today is greatly enhanced by the $A_1$ resonance, 
especially in the region $m_{A_1}/2 m_{\tilde{\chi}^0_1} \gtrsim 0.98$. However, the relic density is  
relatively large in the resonant region $m_{A_1}/2 m_{\tilde{\chi}^0_1} \to 1$, which is mainly because of 
the reduced Breit-Wigner factor due to the large DM energy at freezing out.

From the upper right panel of Fig.~\ref{s1num}, we conclude that the annihilation of DM is still dominated 
by the $\kappa$ coupling for $A_1 \tilde{\chi}^0_1 \tilde{\chi}^0_1$, even though the $\kappa$ is preferably 
more than one order of magnitude smaller than $\lambda$
\begin{align}
\lambda C_{\tilde{\chi}^0_1, d} C_{\tilde{\chi}^0_1, d} C_{A_1, s} \ll \kappa C^2_{\tilde{\chi}^0_1, s} C_{A_1, s}~.~
\end{align}
From the figure, we also find that the coupling between the $A_1$ and the bottom quarks is indeed suppressed 
by it small $H_d$ component.

On the other hand, the invisible decay of the SM-like Higgs boson $H_{\text{SM}}$ 
is dominated by the large $\lambda$ coupling, as shown in the lower left panel of Fig.~\ref{s1num}. 
Thus, we need the smallness of $\tilde{H}_d$ component in $\tilde{\chi}^0_1$ to suppress the coupling 
between the dark matter and SM-like Higgs boson. Note that because of Eq.~(\ref{s1dmmass}), DM mass will
decrease when increasing $\lambda$. The smaller Higgs invisible decay branching  
ratio usually means the heavier dark matter mass.

\begin{figure}[htb]
  \centering
  \includegraphics[width=0.48\textwidth]{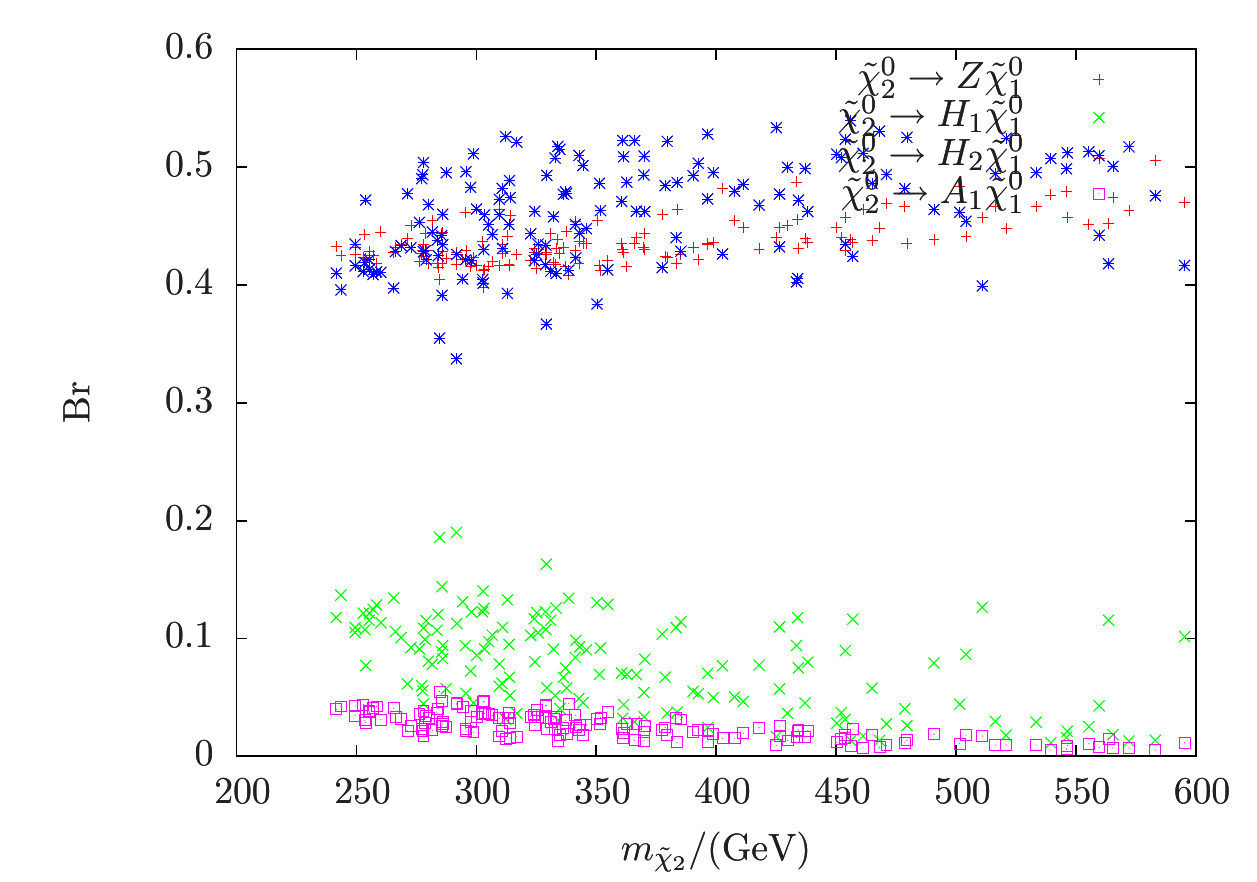}
   \includegraphics[width=0.48\textwidth]{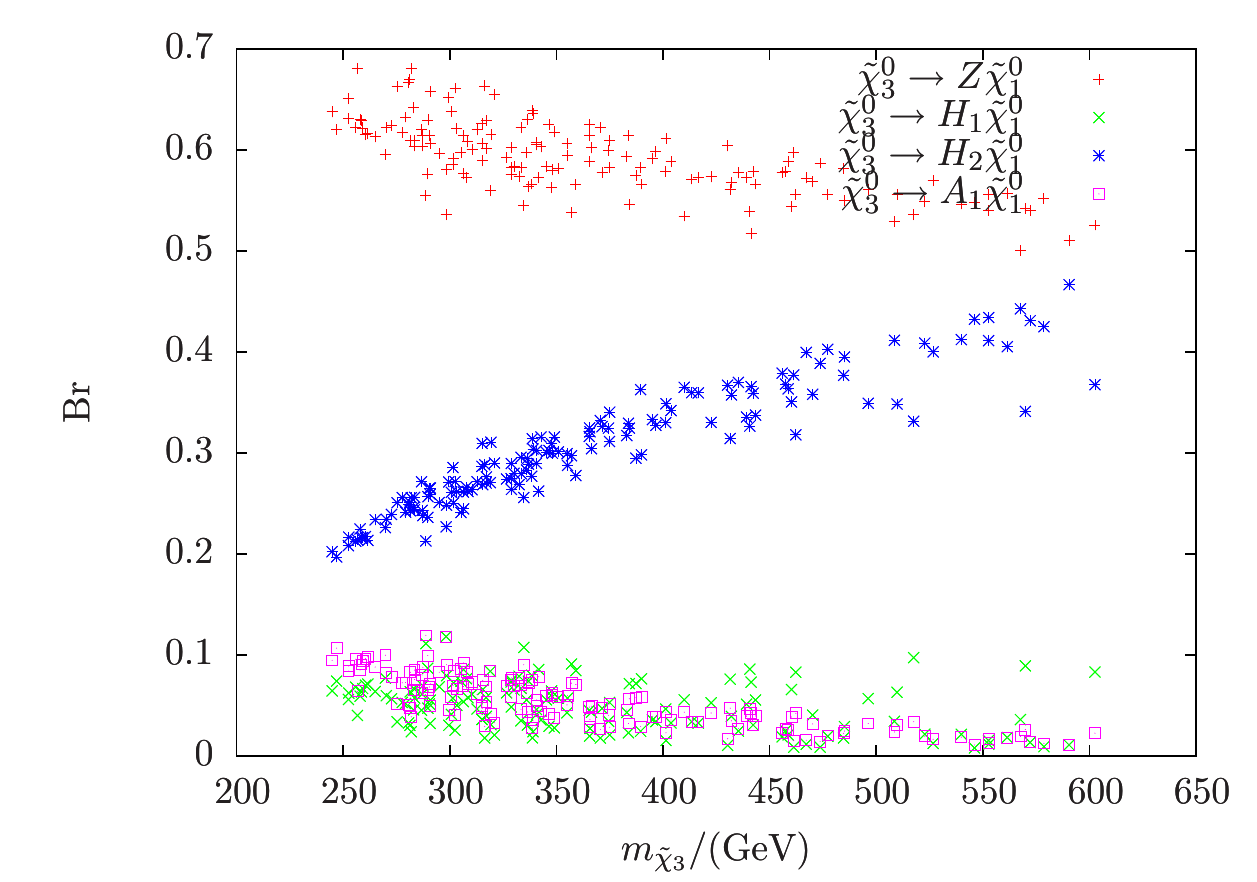}
  \caption{\label{decays1} All the points in figures satisfy the constraints that mentioned 
in the Section~\ref{NMSSM}.  Left: Decay branching ratios for $\tilde{\chi}^0_2$. 
Right: Decay branching ratios for $\tilde{\chi}^0_3$.}
  \label{n23decay}
\end{figure}

As we have emphasized before, the Higgsino component in the DM is very important to achieve 
the correct DM relic density. That is why one usually needs relatively small $\mu$ and 
large tan$\beta$ in this scenario. As shown in the lower right panel of Fig.~\ref{s1num}, 
we find $\mu$ preferably lies in the range of [200,~600] GeV, which gives us a handle to 
discovery this scenario at the LHC, by searching for relatively light Higgsinos. We will present
more extensive study in the next Section.
Fig.~\ref{n23decay} shows Higgsino-like neutralino decay branching ratios, in which both $\tilde{\chi}^0_2$ 
and $\tilde{\chi}^0_3$ have very large decay branching ratios to $Z \tilde{\chi}^0_1$ 
and $H_{\text{SM}} \tilde{\chi}^0_1$. And their decays to lighter Higgs bosons, which usually have branching ratios smaller than 10\%, are 
suppressed by the large singlet component.

\begin{figure}[htb]
  \centering
  \includegraphics[width=0.7\textwidth]{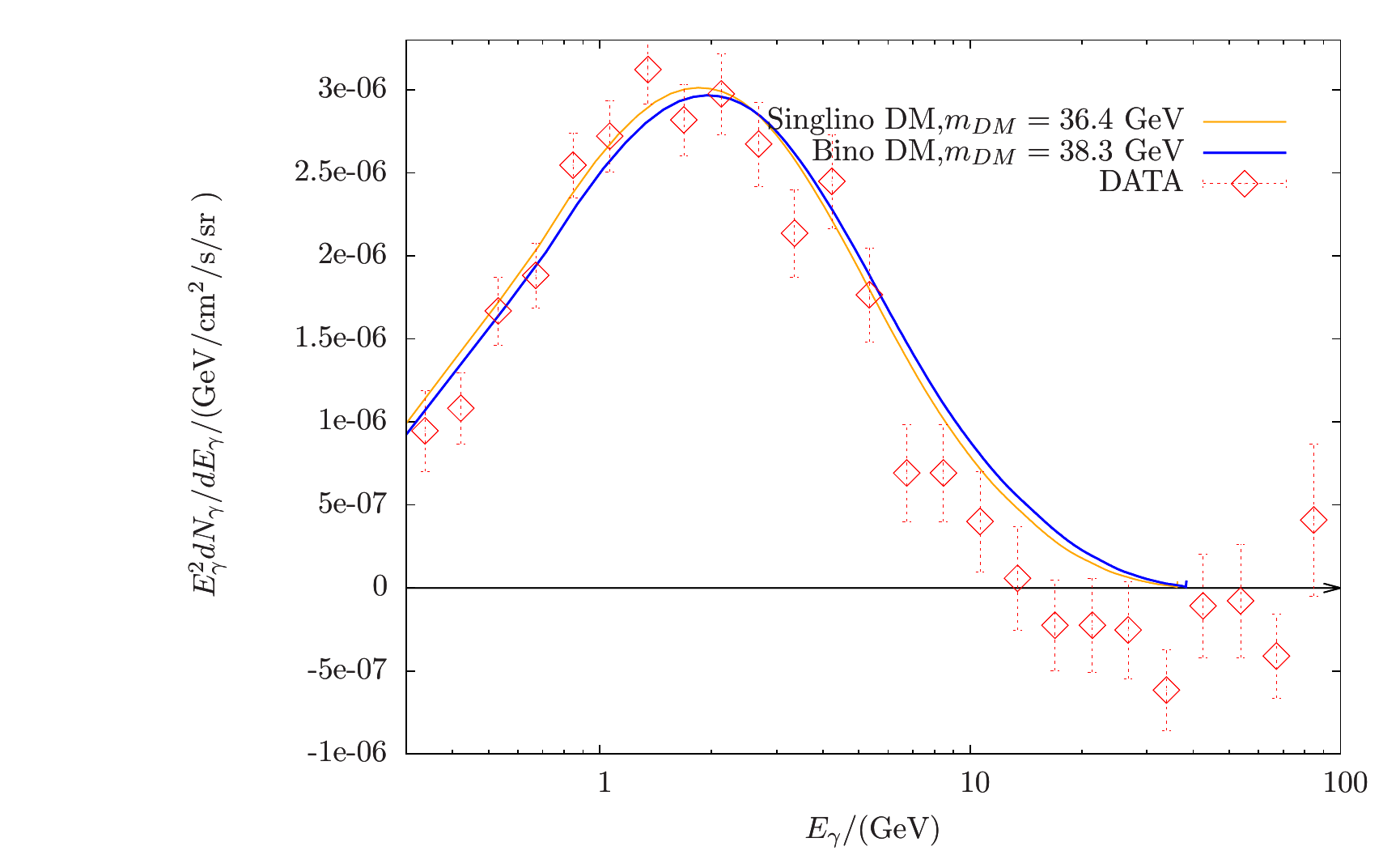}
  \caption{The gamma-ray spectra of two benchmark points in Tables~\ref{s1bp} and \ref{binobmp}.  
The generalised NFW halo profile with an inner slope of $\gamma =1.26$ is chosen. 
And the angle distance from the Galactic Center is $5^\circ$.  }
  \label{s1gamma}
\end{figure}

In order to have a better understanding of this scenario,  we present a benchmark point  in Table~\ref{s1bp}.
 \begin{table}[htb]
 \begin{center}
 \scalebox{0.8}{
  \begin{tabular}{c|c|c|c|c|c} \hline \hline
  $\lambda$ & $\kappa$ & tan$\beta$ & $\mu$ & $A_\lambda$ & $A_\kappa$   \\ \hline
   0.557 & 0.033 & 25.72 & 313.2 & 8029.5 & 9.51 \\ \hline \hline
$A_t$ & $m_{A_1}$&  $m_{H_1}$ & $m_{H_2}$ & $m_{\tilde{\chi}^0_1}$ & $m_{\tilde{\chi}^0_2}$    \\ \hline
3686.4 & 72.1 & 65.3 & 124.9 & 36.4 & 332.1 \\ \hline \hline
$m_{\tilde{\chi}^\pm_1}$ & $\frac{\tilde{\chi}^0_1 \tilde{\chi}^0_1 \to b \bar{b}}{\tilde{\chi}^0_1 \tilde{\chi}^0_1 \to e^+ e^-}|_{\text{\tiny{Freeze Out}}}$ & $\Omega h^2$ &    $\langle \sigma v \rangle|_{v \to 0}(\text{cm}^3/\text{s}) $ & $\sigma_{\text{SI}} (\text{cm}^2)$ & Br$(H_2 \to \tilde{\chi}^0_1 \tilde{\chi}^0_1)$   \\ \hline
319.6 & 5.5 & 0.11  & $2.25 \times 10^{-26}$  & $6.34 \times 10^{-50}$ & 3.9\%   \\ \hline \hline
 Br$(\tilde{\chi}^0_2 \to Z \tilde{\chi}^0_1)$ &  Br$(\tilde{\chi}^0_2 \to H_2 \tilde{\chi}^0_1)$ &  Br$(\tilde{\chi}^0_2 \to H_1 \tilde{\chi}^0_1)$ &  Br$(\tilde{\chi}^0_3 \to Z \tilde{\chi}^0_1)$ &  Br$(\tilde{\chi}^0_3 \to H_2 \tilde{\chi}^0_1)$ &  Br$(\tilde{\chi}^0_3 \to H_1 \tilde{\chi}^0_1)$ \\ \hline
 41.8\% & 46.3 \% & 8.8\% & 59.7\% & 28.7\% & 5.4\% \\ \hline
  \end{tabular}}
  \caption{\label{s1bp} The benchmark point for $s$-Channel A1-resonant annihilation, which satisfies all the constraints mentioned 
in the text and whose other soft parameters are given in Eq.~(\ref{nmssmp1}). All mass parameters are in units of GeV. }
 \end{center}
\end{table}
From this table, the ratios of the DM annihilation into $b \bar{b}$ and $e^+ e^-$ at freeze out are
 quite similar to the decay branching ratios of $Z$ boson. So we conclude that the DM annihilations 
are indeed dominated by the $Z$ boson exchange in the early universe, while in the present universe the $Z$ boson contributions 
suffer from both $p$-wave suppression and from being away from resonance pole.
For this benchmark point, 90\% of the DM annihilates into $b{\bar b}$. Following the same analysis 
as in Ref.~\cite{Daylan:2014rsa}, we calculate the gamma ray spectrum for this benchmark point 
by using micrOMEGAs, which is given in Fig.~\ref{s1gamma}. Thus, it can fit the observed GCE very well.

Finally, let us briefly discuss the scenario where the DM is Bino dominant. 
In this case there is no constraint on the singlino mass so that $\kappa$ becomes a free parameter
and so there is not much difference between this and the singlino LSP scenario. In order to annihilate Bino-dominated
DM effectively,  we still need the $A_1$ resonant enhancement. A relatively large Higgsino component is needed to 
realize the correct DM relic density as well.
As a result, except for the singlet state, the rest of the model properties are quite similar to 
the above singlino LSP scenario. We give a benchmark point in Table~\ref{binobmp}.
For this benchmark point, because the singlet is heavy, the lightest CP-even Higgs boson is SM-like
and its invisible decay to DM is dominated by gauge coupling. The lightest CP-odd Higgs boson 
is singlet-like, with its proper mass $m_{A_1} \sim 2 \times m_{\tilde{\chi}^0_1}$ achieved by tuning $A_\kappa$.
Moreover, the light Higgsino states decay in a manner similar to that in the singlino LSP scenario, 
where the decays are dominated by $Z \tilde{\chi}^0_1$ and $H_{\text{SM}} \tilde{\chi}^0_1 $ final states.

 \begin{table}[h]
 \begin{center}
 \scalebox{0.8}{
  \begin{tabular}{c|c|c|c|c|c} \hline \hline
  $\lambda$ & $\kappa$ & tan$\beta$ & $\mu$ & $A_\lambda$ & $A_\kappa$   \\ \hline
   0.173 & 0.558 & 30.23 & 175.5 & 4220.2 & 0.56 \\ \hline \hline
$A_t$ & $m_{A_1}$&  $m_{H_1}$ & $M_1$ & $m_{\tilde{\chi}^0_1}$ & $m_{\tilde{\chi}^0_2}$    \\ \hline
3234.1 & 76.75 & 124.1 & 41.9 & 38.3 & 184.8 \\ \hline \hline
$m_{\tilde{\chi}^\pm_1}$ & $\Omega h^2$ &    $\langle \sigma v \rangle|_{v \to 0}(\text{cm}^3/\text{s}) $ & $\sigma_{\text{SI}} (\text{cm}^2)$ & Br$(H_1 \to \tilde{\chi}^0_1 \tilde{\chi}^0_1)$ & Br$(A_1 \to b \bar{b})$  \\ \hline
179.7 & 0.118  & $2.38 \times 10^{-26}$  & $3.84 \times 10^{-46}$ & 15\% & 90\%  \\ \hline \hline
 Br$(\tilde{\chi}^0_2 \to Z \tilde{\chi}^0_1)$ &  Br$(\tilde{\chi}^0_2 \to H_1 \tilde{\chi}^0_1)$ &  Br$(\tilde{\chi}^0_2 \to A_1 \tilde{\chi}^0_1)$ &  Br$(\tilde{\chi}^0_3 \to Z \tilde{\chi}^0_1)$ &  Br$(\tilde{\chi}^0_3 \to H_1 \tilde{\chi}^0_1)$ &  Br$(\tilde{\chi}^0_3 \to A_1 \tilde{\chi}^0_1)$ \\ \hline
 52.7\% &  45.8\% & 1.5\% & 90\% & 8.2\% & 1.8\% \\ \hline
  \end{tabular}}
  \caption{The benchmark point with Bino-like DM, which satisfies all the constraints mentioned in the text and 
whose other soft parameters are given in Eq.~(\ref{nmssmp1}). All mass parameters are in units of GeV. \label{binobmp}}
 \end{center}
\end{table}

\subsection{Hidden Sector Dark Matter in the NMSSM}
\label{NMSSMs2}
In this subsection, we will briefly comment on another possible explanation of GCE within the $Z_3$ NMSSM, where DM
with mass $m_{\tilde{\chi}^0_1}\sim 67$  GeV 
can predominantly annihilate into $H_1$ and $A_1$.
Both $H_1$ and $A_1$ are singlet-like while having small fractions of $H_d$ component. 
Then $H_1$ and $A_1$ mainly decay into $b \bar{b}$  because of the relatively large bottom quark Yukawa coupling.
In this case, we can not employ the resonant enhancement. So, DM can only be singlet-like and a very large coupling of the singlet state $\kappa$ is required. 
The singlet mass  $m_{\chi}\simeq 2\frac{\kappa}{\lambda}\mu$ implies 
\begin{align}
\frac{\kappa}{\lambda}=\frac{m_{\widetilde{\chi}_1^0}}{2\mu_{\text{eff}}}\lesssim \frac{67}{200}~.~
\label{dm70}
\end{align}

The DM direct detection experiments also give strong constraints on the DM-nucleon scattering cross section. 
The main contributions to direct detection come from $H_1$ and $H_2$ mediated
 $t$-channel processes:
\begin{align}
\sigma_{SI}\simeq\frac{\kappa^2\mu_{\widetilde{\chi}^0_1 N}^2m_{N}^2f_N^2}{4\pi v^2}(\frac{C_{H1u}}{m_{H_1}^2}+\frac{C_{H2s}}{m_{H_2}^2})^2~,~\,
\end{align}
where $f_N=\sum_{q=u,d,s}f_{T_q}+\frac{2}{9}f_{T_G}\simeq 0.348\pm0.015$~\cite{Cline:2013gha}, $C_{H1u}$ and $C_{H2s}$ are
 the $H_u$ component in $H_1$ and  singlet component in $H_2$ respectively. To obtain a small spin-independent cross section,
 $\sigma_{SI}$,  it is necessary to have a quite small mixing between $H^0_u$ and $S$ due to 
\begin{align}
C_{H1 u}^2\simeq \sigma_{SI}\frac{4\pi v^2 m_{H_1}^4}{0.1^2f_N^2}(\frac{0.1}{\kappa})^2<0.004 \, .
\end{align}
To suppress the singlet component in the SM-like $H_2$, we use the condition 
$A_{\lambda} = 2 \mu/\sin 2 \beta - 2 \kappa \mu /\lambda$.
Moreover, from Eq.~(\ref{dm70}) and the Landau pole condition $\lambda^2+\kappa^2 \lesssim 0.5$,  we have
\begin{align}
| \kappa | \lesssim \sqrt{\frac{1}{2}( \frac{m^2_{\tilde{\chi}^0_1}}{m^2_{\tilde{\chi}^0_1} + 4 \mu^2} )  }~.~ \label{ck}
\end{align}

On the other hand, the cross section of dark matter annihilation into $H_1$ and $A_1$ is expected to be
\begin{align}
\langle\sigma v\rangle \simeq \frac{\kappa^4}{4\pi m_{\widetilde{\chi}^0_1}^2} \frac{|\overrightarrow{P_3}|}{m_{\widetilde{\chi}^0_1}}
                              (\frac{4m_{\widetilde{\chi}^0_1}^2+m_{A_1}^2-m_{H_1}^2}{4m_{\widetilde{\chi}^0_1}^2-m_{A_1}^2-m_{H_1}^2}-\frac{m_{\widetilde{\chi}^0_1}^2-A_{\kappa}m_{\widetilde{\chi}^0_1}}{4m_{\widetilde{\chi}^0_1}^2-m_{A_1}^2})^2~,~
\label{dm70sigmaV}
\end{align}
in which $|\overrightarrow{P_3}|=\sqrt{E_{A_1}^2-m_{A_1}}$, and $E_{A_1}=\sqrt{\frac{4m_{\chi}^2+m_{A_1}^2-m_{H_1}^2}{4m_{\chi}}}$. 
As a result, the annihilation cross section is highly constrained by Eq.~(\ref{ck}).

Following the same strategy as above, we survey the parameter space in the low energy NMSSM.
We choose the same irrelevant soft terms as Eq.~(\ref{nmssmp1}) and the ranges for the DM related parameters as follows
\begin{align}
\lambda: [0.35, 0.65], ~ \kappa: -[50,100]\times \frac{\lambda}{2 \mu}, ~ \tan \beta: [1,40] \nonumber \\
\mu: [100,500] ~\text{GeV}, ~ A_{\kappa}: [-100,100] ~\text{GeV} \\
A_{\lambda} = 2 \mu/\sin 2 \beta - 2 \kappa \mu /\lambda, ~ A_t = \mu/\tan \beta + \sqrt{6}\times1500 ~,~ \nonumber
\end{align}
where a negative $\kappa$ is taken to reduce the SM-like Higgs invisible decay into $A_1 A_1$.

\begin{figure}[htb]
\includegraphics[width=3.0in]{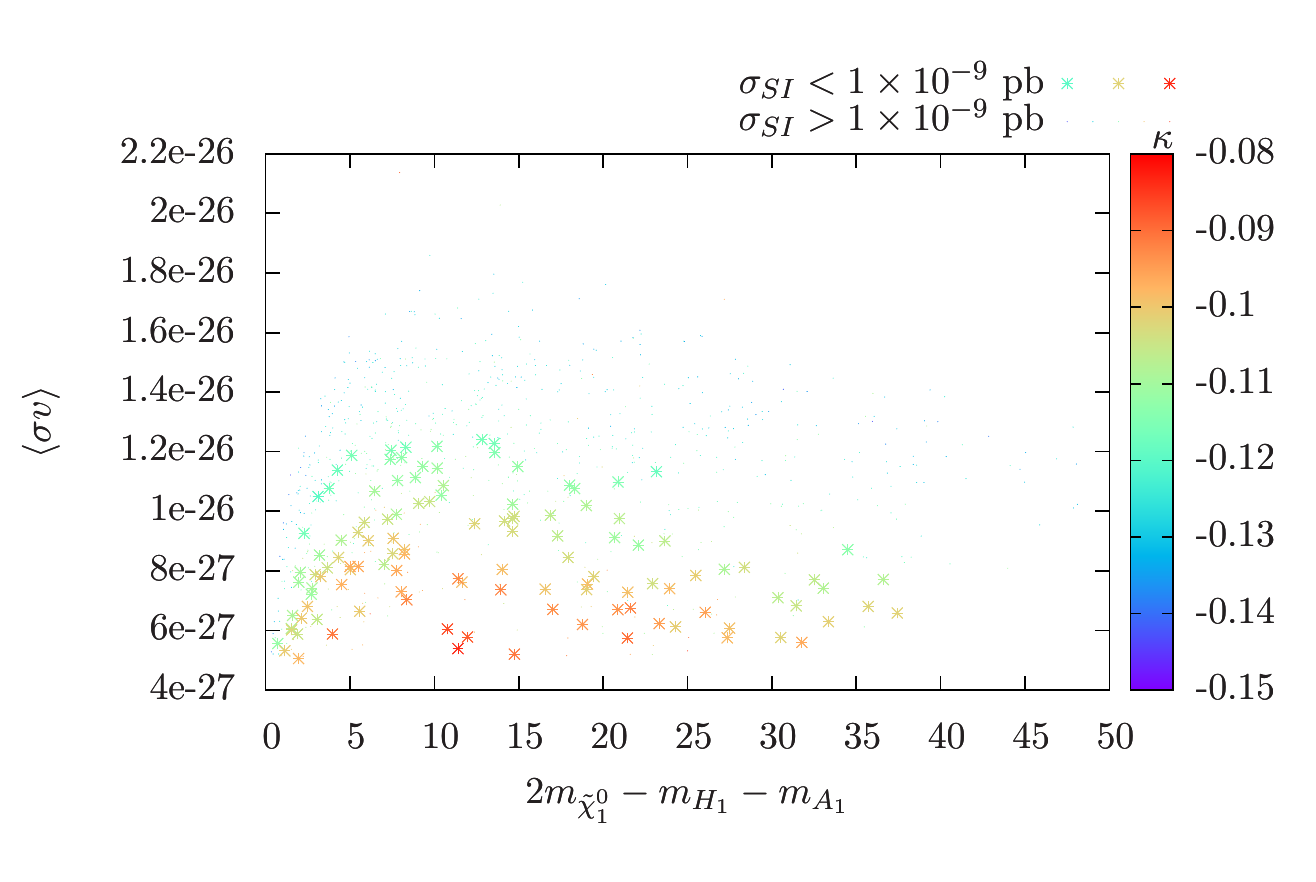}
\includegraphics[width=3.0in]{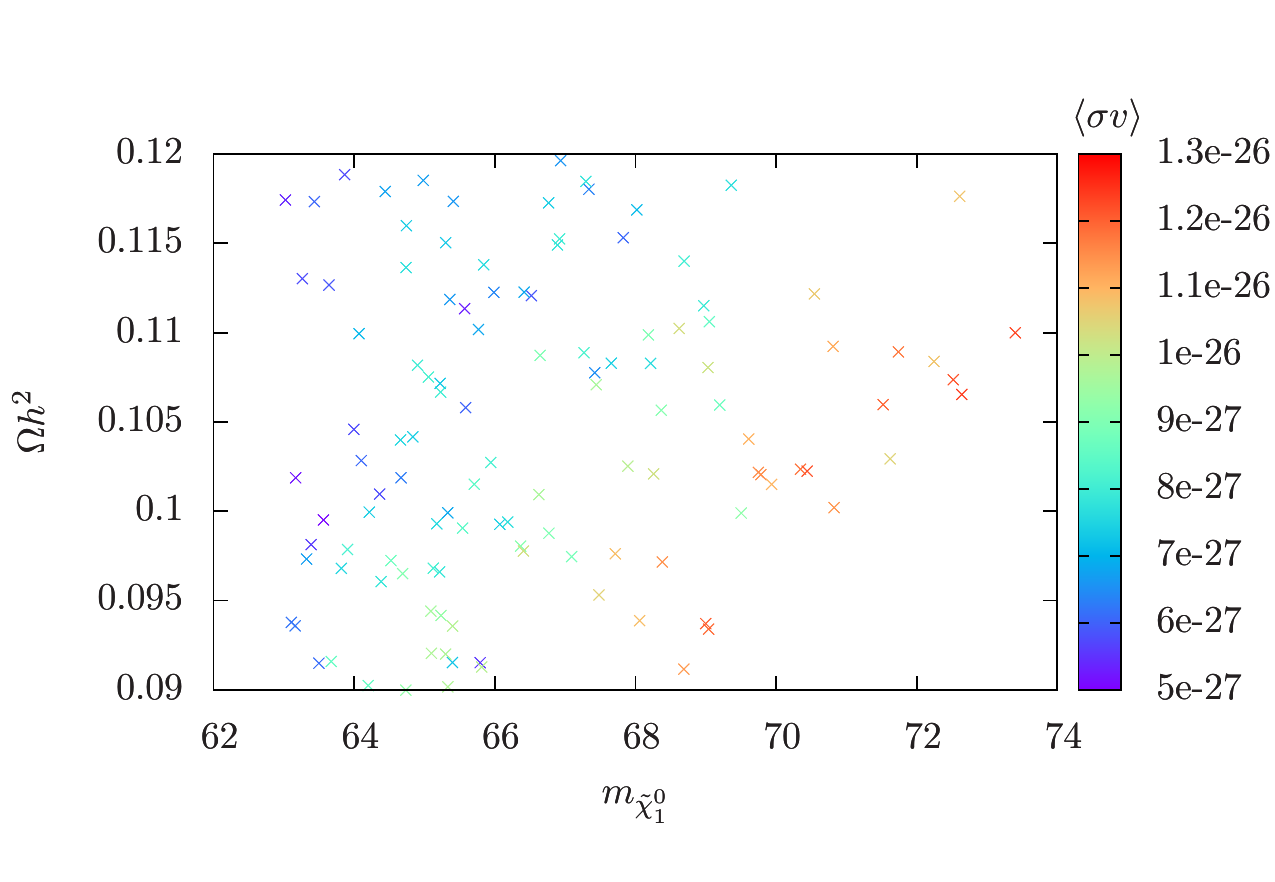}
\caption{\label{hiddenlux}
Left: The recent direct detection LUX experiment gives a strong constraint. Because the larger $\kappa$ means larger $\sigma_{SI}$, 
 it is difficult to get a big $\langle \sigma v\rangle$ and $\langle \sigma v\rangle>1.5\times10^{-26}
~\text{cm}^3/s$ has been excluded by the LUX results. Right: All the points satisfy the constraints mentioned 
in the last two figures, so we can find an upper limit on $\langle \sigma v\rangle$.
}
\end{figure}

As shown in the left panel of Fig.~\ref{hiddenlux}, the LUX experiment constrains $|\kappa|$ to be no larger than $0.12$. 
This will lead to the DM annihilation cross section smaller than $\sim 1.2 \times 10^{-26}\text{cm}^3/s$. From the right panel of the figure, we conclude that there is no great tension 
between the DM density and DM annihilation rate. However, a heavier DM tends to have 
a relatively large annihilation rate because of large $|\kappa|$.

We present a benchmark point for this scenario in Table~\ref{s2bp}. This case usually has very small $\mu$ 
and relatively large $|\kappa|$. The spin-independent DM-nucleon cross section is close to the exclusion limit, 
while the annihilation cross section only reaches $1.22 \times 10^{-26} \text{cm}^3/s$.

 \begin{table}[htb]
 \begin{center}
 \scalebox{0.8}{
  \begin{tabular}{c|c|c|c|c|c} \hline \hline
  $\lambda$ & $\kappa$ & tan$\beta$ & $\mu$ & $A_\lambda$ & $A_\kappa$   \\ \hline
   0.374 & -0.113 & 26.09 & 143.82 & 3846.26 & 39.716 \\ \hline \hline
$A_t$ & $m_{A_1}$&  $m_{H_1}$ & $m_{H_2}$ & $m_{\tilde{\chi}^0_1}$ & $m_{\tilde{\chi}^0_2}$    \\ \hline
3679.7 & 43.9 & 86.9 & 125.9 & 70.43 & 153.3 \\ \hline \hline
$m_{\tilde{\chi}^\pm_1}$ & $\Omega h^2$ &    $\langle \sigma v \rangle|_{v \to 0}(\text{cm}^3/\text{s}) $ & $\sigma_{\text{SI}} (\text{cm}^2)$ & Br$(H_2 \to A_1 A_1)$ & Br$(H_1 \to b \bar{b})$  \\ \hline
147.3 & 0.1026  & $1.22 \times 10^{-26}$  & $9.43 \times 10^{-46}$ & 4.5\% & 89.6\%  \\ \hline \hline
 Br$(\tilde{\chi}^0_2 \to Z \tilde{\chi}^0_1)$ &  Br$(\tilde{\chi}^0_2 \to H_2 \tilde{\chi}^0_1)$ &  Br$(\tilde{\chi}^0_2 \to A_1 \tilde{\chi}^0_1)$ &  Br$(\tilde{\chi}^0_3 \to Z \tilde{\chi}^0_1)$ &  Br$(\tilde{\chi}^0_3 \to H_2 \tilde{\chi}^0_1)$ &  Br$(\tilde{\chi}^0_3 \to H_1 \tilde{\chi}^0_1)$ \\ \hline
 0\% & 0 \% & 96.3\% & 62.3\% & 0\% & 29.96\% \\ \hline
  \end{tabular}}
  \caption{\label{s2bp} The benchmark point for hidden sector DM scenario. which satisfies all the constraints mentioned in the text.
Also, the other soft parameters are given in Eq.~(\ref{nmssmp1}). All mass parameters are in units of GeV. }
 \end{center}
\end{table}

In short, the GCE in this scenario is constrained by the Landau pole condition and the DM direct search limits. 
It is difficult to have the DM annihilation cross section as large as expected, $2.2 \times 10^{-26} ~\text{cm}^3/s$. 
Nonetheless, the situation changes if we take into account the large uncertainty in background analysis~\cite{Agrawal:2014oha,Calore:2014nla},
where the allowed annihilation rate can vary in a large range $\sim [0.4,4.5] \times 10^{-26}~\text{cm}^3/s$. 
We will not consider this case further in this work. 

\section{Discovery Potential at the LHC}
\label{LHC}

We have shown that the s-channel $A_1$ mediated annihilation with singlino dominant DM is the most promising scenario and
so we will restrict our attention to its discovery potential at the LHC. 
For this scenario, we have a light Higgs sector and a light singlino and Higgsino. 
The pseudoscalar, even though relatively light, is nearly pure singlet as shown in Fig.~\ref{s1num}. So, it will be very difficult to produce it at the hadron collider. Moreover, we find it is very difficult to produce the pseudoscalar from the SM-like Higgs mediated processes and neutralino decays as well. 
The deviation of the coupling and invisible decay of the SM-like Higgs boson may also provide a smoking gun for this scenario. 
On the one hand, both the measurement accuracy of couplings that can be reached at the LHC~\cite{Peskin:2012we} and the constraint on Higgs invisible decay~\cite{Aad:2014iia} are very weak. On the other hand, and more importantly, the Higgs couplings and its invisible decay width vary in a large range in our scan, with many of the models having very SM-like couplings and suppressed invisible decay width, as indicated in Fig.~\ref{s1num}. So, those signatures should not be able to provide very promising signals 
in the full parameter space of the scenario.
However, in the electroweakino sector, there are  two remarkable signatures for the viable NMSSM with GCE.
The first one is that the DM particle has a relatively large Higgsino component and a moderately large coupling to the $Z$ boson 
which guarantee the correct DM relic density. So, a natural thought will be the direct production of DM pair at the LHC 
through $s$-channel $Z$ boson exchange. The second signature is the existence of a relatively light Higgsino  
together with  $\sim 35$ GeV DM, whose dominant decay modes are
\begin{align}
\tilde{H}^\pm &\to W^\pm \tilde{\chi}^0_1, \\
\tilde{H}^0_{1,2}& \to Z \tilde{\chi}^0_1/ H \tilde{\chi}^0_1~.~
\end{align}
The decay branching ratios are given in Fig.~\ref{decays1}.
We shall discuss  each signature in detail in the following.

\subsection{Dark Matter Production versus Mono-jet}

If only one DM pair is produced at a hadron collider, it will not leave any information inside the detector. 
One way to probe this process is through a hard initial state radiation (ISR) jet, which is recoiling against the DM pair. 
As result, the signal presents a large unbalanced transverse energy. At the LHC, the CMS Collaboration 
has carried out a mono-jet search~\cite{Aad:2014nra} with a data set of 20 fb$^{-1}$. Their analysis shows 
that the current stage of the LHC is only sensitive to signals with $p_T(\text{ISR}) > 280, ~340, ~450$ GeV that have 
cross section greater than about $100, ~30, ~10$ fb, respectively.

\begin{figure}[htb]
\includegraphics[width=0.48\textwidth]{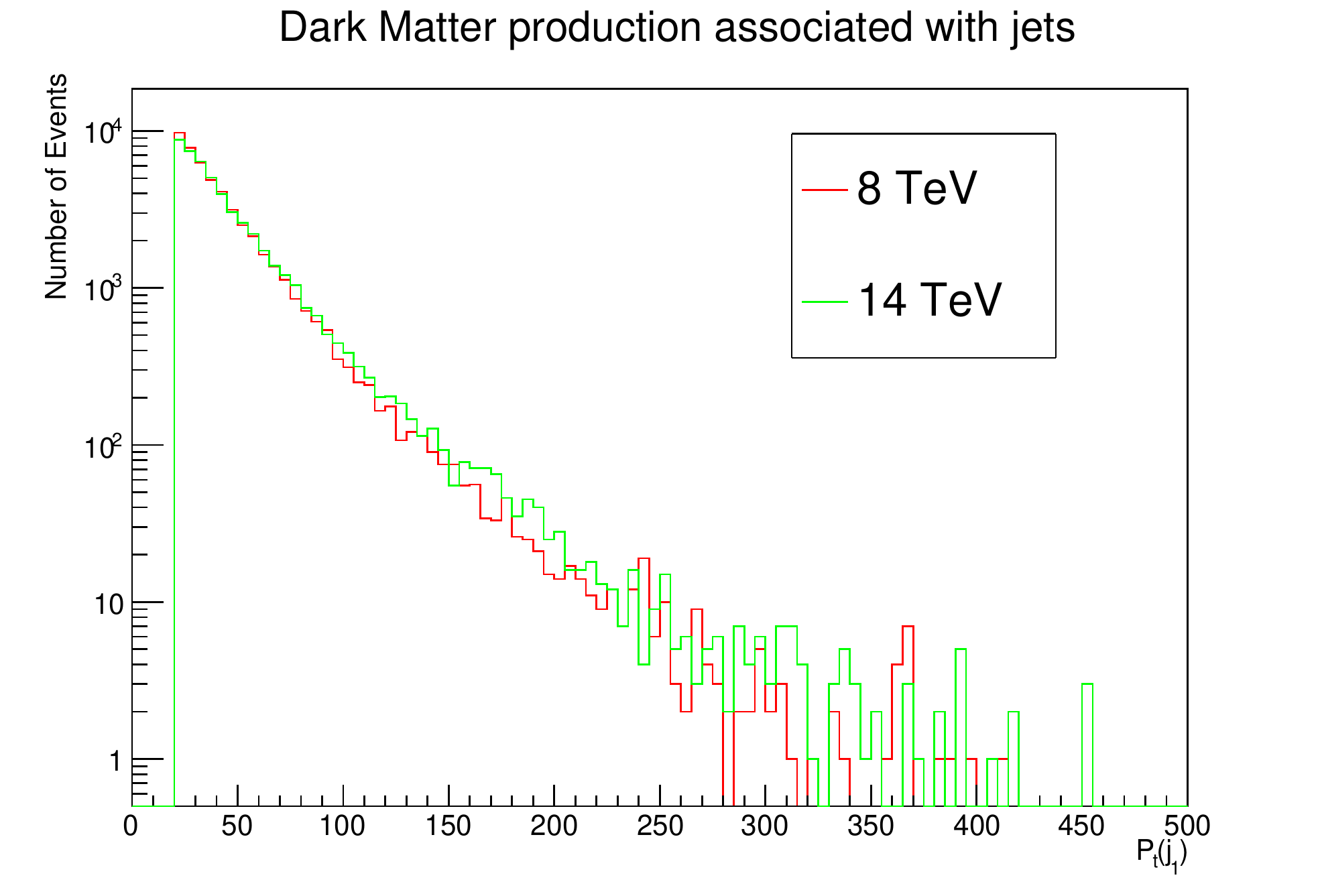}
\includegraphics[width=0.48\textwidth]{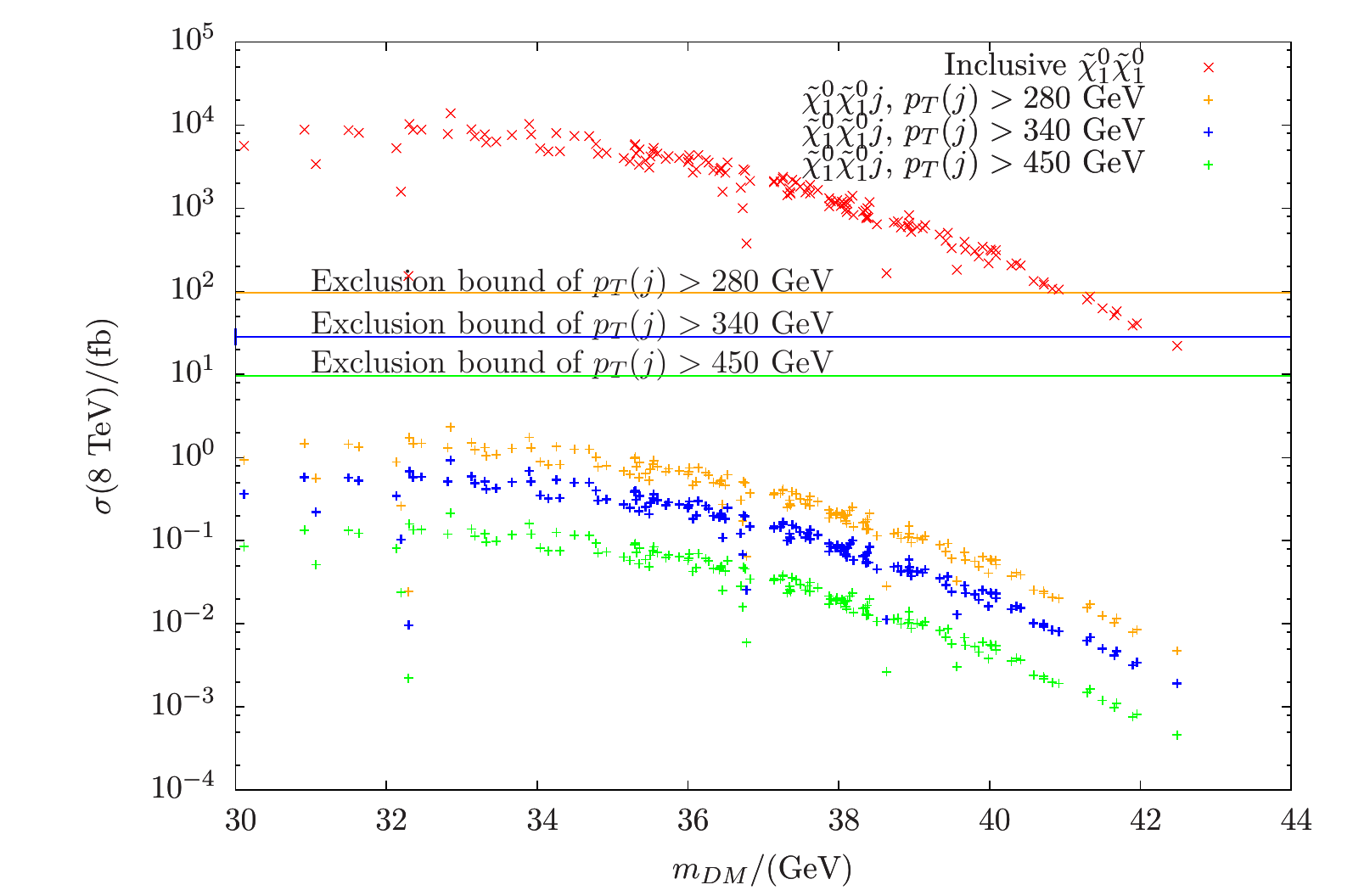}
\caption{\label{monofig} Left: the $p_T$ spectrum of the leading ISR jet for $35.5$ GeV DM pair production at LHC. 
Right: the cross section in the models for the GCE with corresponding cuts on the leading ISR jet as indicated 
in the figure. The horizontal lines are the experimental bounds on cross sections for three categories of the models. }
\end{figure}

For our process, even though the direct production is relatively large ($\sim$10 pb) because of 
the small DM mass, the production rate of the mono-jet signal is quite low. 
This is because, as has been studied in Ref.~\cite{Alwall:2008qv}, the spectrum of the radiated jet mainly depends on 
the mass scale of the final states. In our case, because the DM mass is very small ($\sim 35$ GeV)
the production of an energetic ISR jet is suppressed. From the left panel of Fig.~\ref{monofig}, we conclude 
that the $p_T$ spectrum of the ISR jet drops very quickly and only $\sim 1/10^{-4}$ of the events have $p_T(\text{ISR}) > 250$ GeV. 
The situation will improve very much in the next run of the LHC at 14 TeV.
We show the cross section for the allowed models with different cuts on the leading ISR jet and 
the corresponding CMS exclusion bounds in the right panel of Fig.~\ref{monofig}.

The production cross section drops by about four orders of magnitude after we  require
 the leading ISR to have $p_T> 280$ GeV. So the signal is around two orders of magnitude smaller 
than the current bound. The sensitivity will not become better for harder cuts on the leading ISR jet.
So, we may conclude that it will be very difficult to probe the GCE NMSSM through mono-jet signatures at the LHC, even at the 14 TeV upgraded
LHC.

\subsection{Searching the Di-Higgs Final State}

The detection of Higgs boson pair production is not only important for the potential to reveal new physics 
but also important for testing the SM itself.
A signature of our models is that the Higgs particles can be pair produced by neutral Higgsino pair production with
subsequently decay into Higgs and DM with large branching ratios as shown in Fig.~\ref{decays1}.

\begin{figure}[htb]
\includegraphics[width=0.48\textwidth]{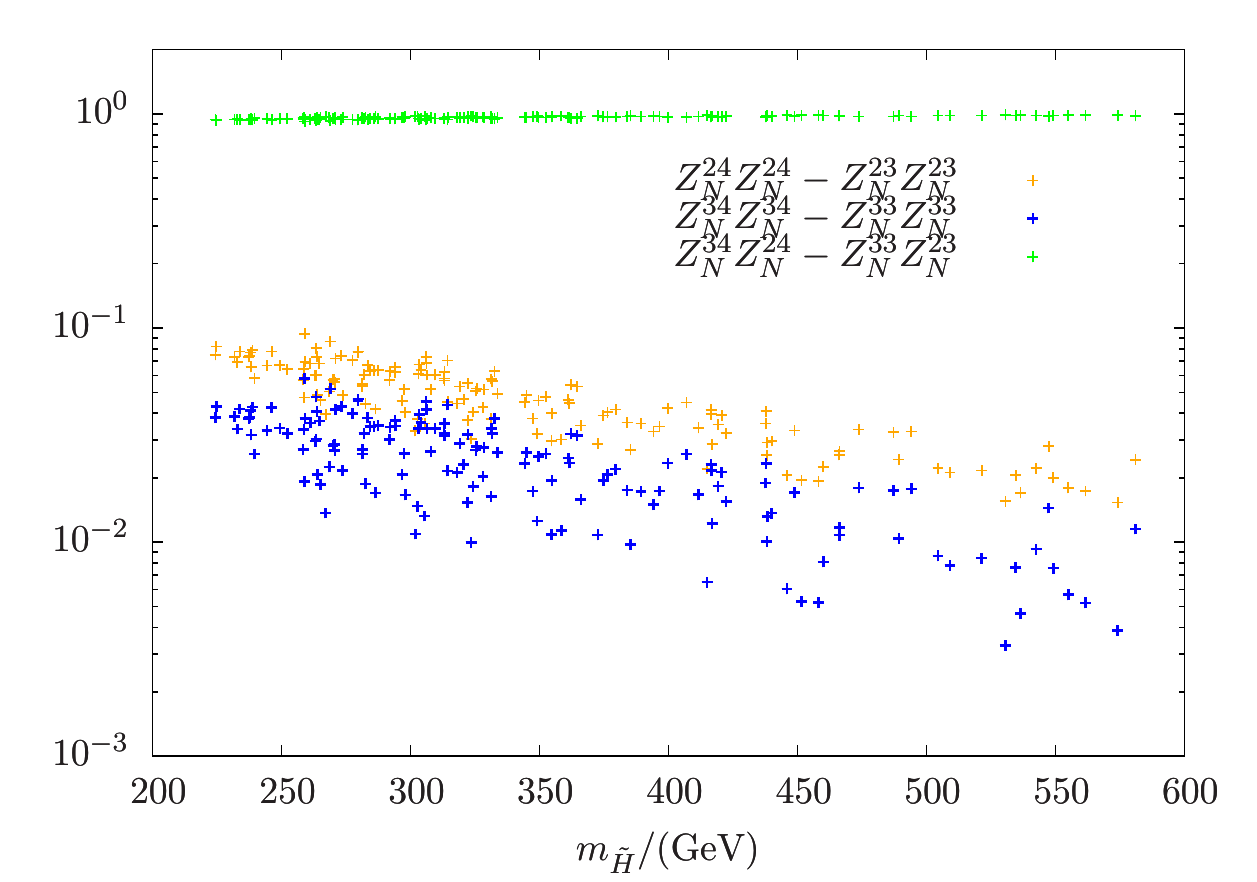}
\includegraphics[width=0.48\textwidth]{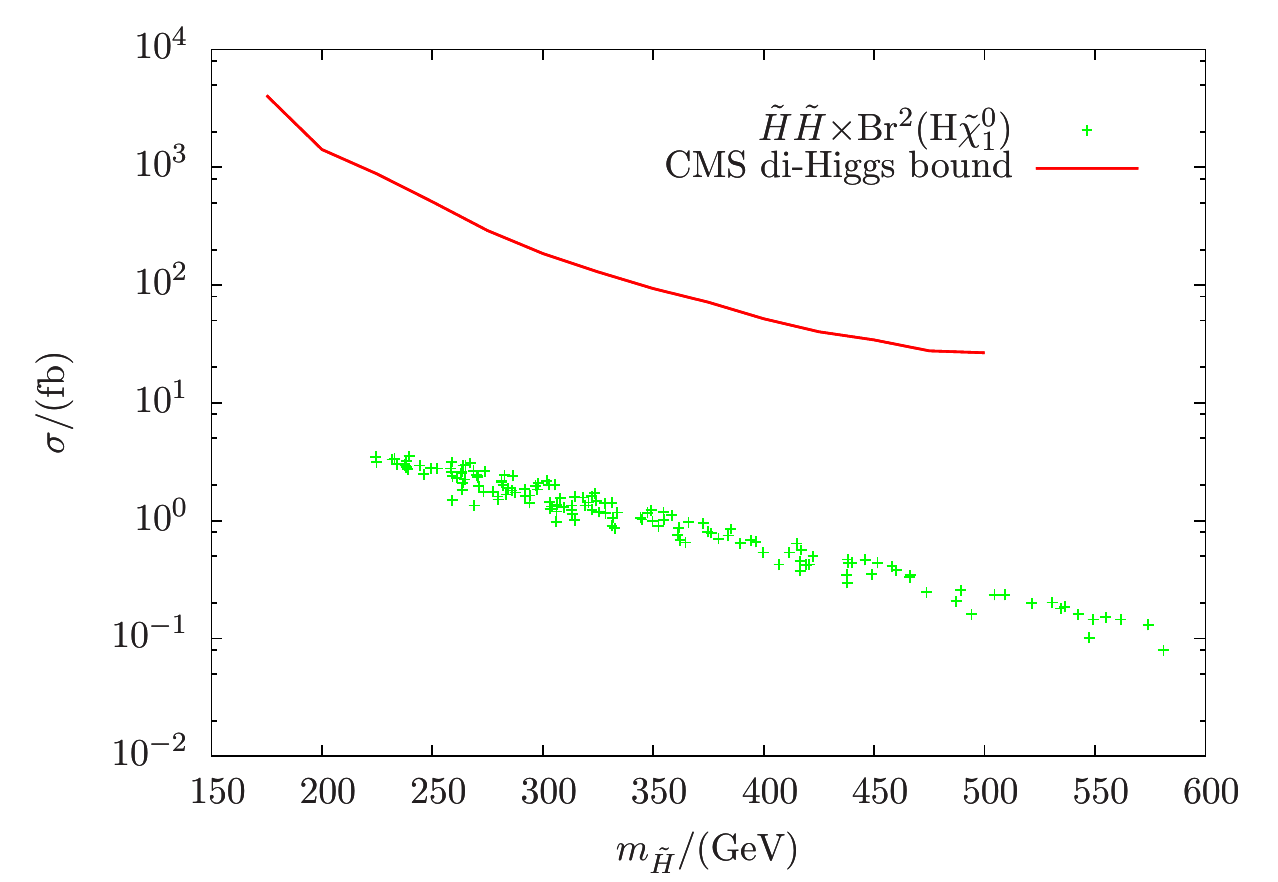}
\caption{\label{hhbound} Left: the mixing factor of $Z$ coupling to each pair of neutral Higgsinos. 
Right: the cross section for the signal models and exclusion bound by the CMS Collaboration.}
\end{figure}

At the LHC, the neutral Higgsino pair is mainly produce through $s$-channel $Z$ boson exchange. As a result, 
the production cross section is dominated by the coupling between the $Z$ boson and Higgsino pair
\begin{align}
Z_\mu \tilde{\chi}^0_i \tilde{\chi}^0_j = \frac{i e}{2 s_W c_W} \gamma^\mu ((Z^*_{i4} Z^*_{j4} - Z^*_{i3} Z^*_{j3} )P_L + (Z^*_{j4} Z^*_{i4} - Z^*_{j3} Z^*_{i3})P_R)~,~\,
\end{align}
where $N_{ij}$ is the neutralino mixing matrix. We show the mixing factor $(Z^*_{i4} Z^*_{j4} - Z^*_{i3} Z^*_{j3})$ for 
the different combinations in the left panel of Fig.~\ref{hhbound}.
From this figure, we know that in our case, the dominant production of neutral Higgsino pair 
is $\tilde{\chi}^0_2 \tilde{\chi}^0_3$, {\it i.e.}, the pair with different masses.

In Ref.~\cite{CMS:2014ica}, the CMS Collaboration has also tried to search for the di-Higgs signal from Higgsino decay. 
Because there is only a small excess in the observed number of events, a very loose bound is obtained as shown by the red line 
in the right panel of Fig.~\ref{hhbound}. The green points in the same figure show the corresponding cross sections 
of our signal processes. Our models are far beyond the reach of a di-Higgs analysis based on the current data base.

\subsection{Electroweakino Searches at 14 TeV LHC}

In this subsection, we will study two traditional but much more sensitive channels
\begin{align}
\tilde{\chi}^0_{2,3} \tilde{\chi}^{\pm}_1 & \to H_{\text{SM}} \tilde{\chi}^0_{1} + W^\pm \tilde{\chi}^0_{1} ~,~\\
\tilde{\chi}^0_{2,3} \tilde{\chi}^{\pm}_1 & \to Z \tilde{\chi}^0_{1} + W^\pm \tilde{\chi}^0_{1} ~.~ 
\end{align}
In fact, these two channels also give rise to the most stringent bounds on Winos at 8 TeV LHC searches. 
For $WH$ final states~\cite{TheATLAScollaboration:2013zia}, the Wino has been excluded for masses up to $\sim 300$ GeV 
for $m_{\tilde{\chi}^0_1} \sim 35$ GeV.
For $WZ$ final states~\cite{Aad:2014vma}, all Wino masses below 420 GeV for $m_{\tilde{\chi}^0_1} \sim 35$ GeV are excluded.
As in our case, the production rates of the chargino and neutralino pairs are suppressed 
due to the mixing between Higgsino states. Moreover, the required final states are suppressed by their decay branching ratios. 
Thus, in order to get improved sensitivities to the NMSSM model with GCE, we have to consider the 14 TeV LHC.

\subsubsection{$WH$ Channel}

The ATLAS analysis in Ref.~\cite{TheATLAScollaboration:2013zia}, where the $W$ boson decays leptonically  
while the Higgs boson decays to $b\bar{b}$, gives the strongest bound for the $WH$ channel at the current stage of LHC. 
Because the ATLAS search had interest in neutralino and chargino mass regions similar to ours, we choose to use those discriminate variables
in their paper to separate the signal and backgrounds. 
We follow their analysis at 8 TeV and attempt to extrapolate to 14 TeV, where some minor changes are applied in order to optimize the search. 
There have been some other studies~\cite{Baer:2012ts,Ghosh:2012mc} of the $WH$ channel at the 14 TeV LHC, which used similar cuts. However, their results are either based on a specific model or on simplified models and we were unable to directly apply their studies in our analysis.  Refs.~\cite{Han:2013kza,Papaefstathiou:2014oja} also studied the $WH$ channel with $H$ decaying to various final states, however the sensitivities of the other final states considered
were not much improved compared to the dominant $b\bar bl^\pm + E^{\text{miss}}_T$ because of their suppressed branching ratios. A very recent ATLAS analysis~\cite{Aad:2015jqa} considered the $\gamma \gamma l$ and same-sign di-lepton final states as well. Their results showed that the strongest expected exclusion limit for heavy electroweakino region ($\sim 200$ GeV) is from $b\bar{b} l^\pm$ final states. Here, our cuts are chosen as below:
\begin{itemize}
\item Exactly one lepton with $p_T>25$ GeV;
\item There should be no more than 3 jets in the event;
\item The leading two jets should be $b$-tagged and their invariant mass is in the range [105,~135] GeV. 
Furthermore, the contransverse mass of the two b-jets 
\begin{align}
m^2_{\text{CT}} = (E^{b1}_T + E^{b2}_T)^2- |\textbf{p}^{b1}_T - \textbf{p}^{b2}_T|^2
\end{align}
is larger than 160 GeV;
\item To remove the $W+$ jets background, the transverse mass of the lepton and missing transverse momentum
\begin{align}
m_T=\sqrt{2 p^{\text{lep}}_T E^{\text{miss}}_T - 2  \textbf{p}^{\text{lep}}_T \cdot \textbf{p}^{\text{miss}}_T }
\end{align}
is required to be larger than 130 GeV since our Higgsinos tend to be heavier than 200 GeV; and
\item We define two signal regions SRA and SRB for different masses of Higgsino which correspond 
to  $E^{\text{miss}}_T > 100$ and $E^{\text{miss}}_T > 245$, respectively. We find 
that the SRB is more sensitive for Higgsino mass greater than 300 GeV.
\end{itemize}

The dominant backgrounds for our analyses are $t\bar{t}$, single top, and $W b \bar{b}$. We generate those backgrounds
 by MadGraph5~\cite{Alwall:2011uj}, where Pythia6~\cite{Sjostrand:2006za} and Delphes 3.1.2~\cite{deFavereau:2013fsa} 
have been used to implement parton shower and detector simulation.
The $t\bar{t}$ is generated up to two additional jets, where the MLM matching adopted in MadGraph5 is used to 
avoid double counting between matrix element and parton shower. All three single top production modes 
($t$-channel, $s$-channel and $tW$ process) are considered in our generation. 
We use the default ALTAS setup for detector simulation.
Their cross sections before and after the cuts for both signal regions are shown in Table~\ref{whbg}.

 \begin{table}[htb]
 \small
 \begin{center}
  \begin{tabular}{c|c|c|c}
   &   $\sigma$(14 TeV)/pb   & SRA/fb & SRB/fb \\ \hline
   $t\bar{t}$ & 877.2 & 0.39  & $2.3 \times 10^{-2}$ \\
   single top & 318.5 & 0.064 & $3.2 \times 10^{-3}$ \\
   $Wb\bar{b}$ & 128 & 0.0096 & $6.4 \times 10^{-4}$ \\
  \end{tabular}
  \caption{\label{whbg} The background cross sections before and after the cuts for each signal region.}
 \end{center}
\end{table}

As for the cut efficiency of signal processes $\epsilon_s$, because we have DM mass $\sim 35$ GeV and Higgsino mass 
in the range [200, 600] GeV, we generate the process of $\tilde{\chi}^0_2 \tilde{\chi}^\pm_1$ pair production with 
subsequent decays $\tilde{\chi}^\pm_1 \to W^\pm \tilde{\chi}^0_1$ and $\tilde{\chi}^0_2 \to H_{\text{SM}} \tilde{\chi}^0_1$  
in Madgraph5 as well. The step size for Higgsino mass is chosen as 20 GeV.

\begin{figure}[htb]
\includegraphics[width=0.48\textwidth]{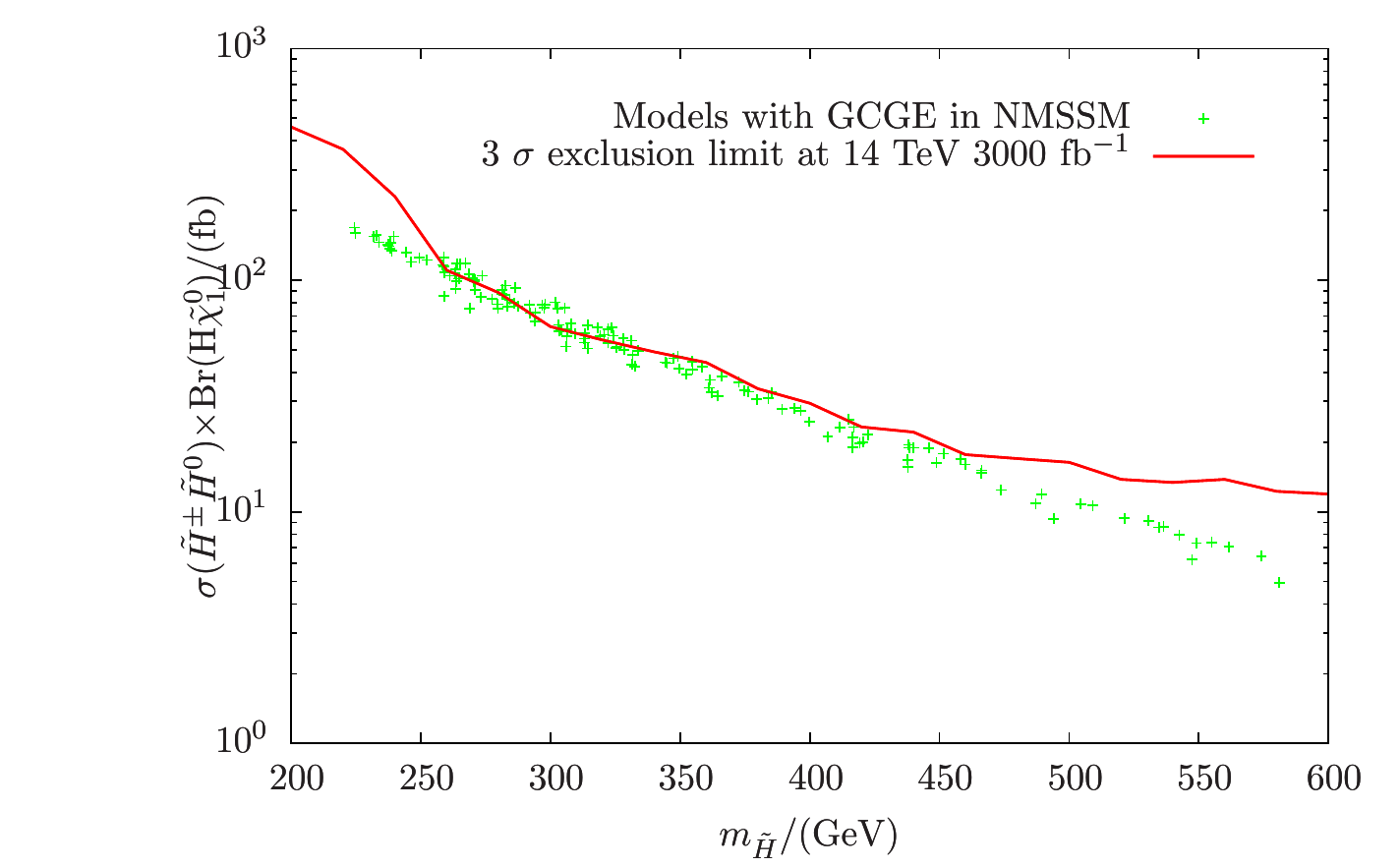}
\includegraphics[width=0.48\textwidth]{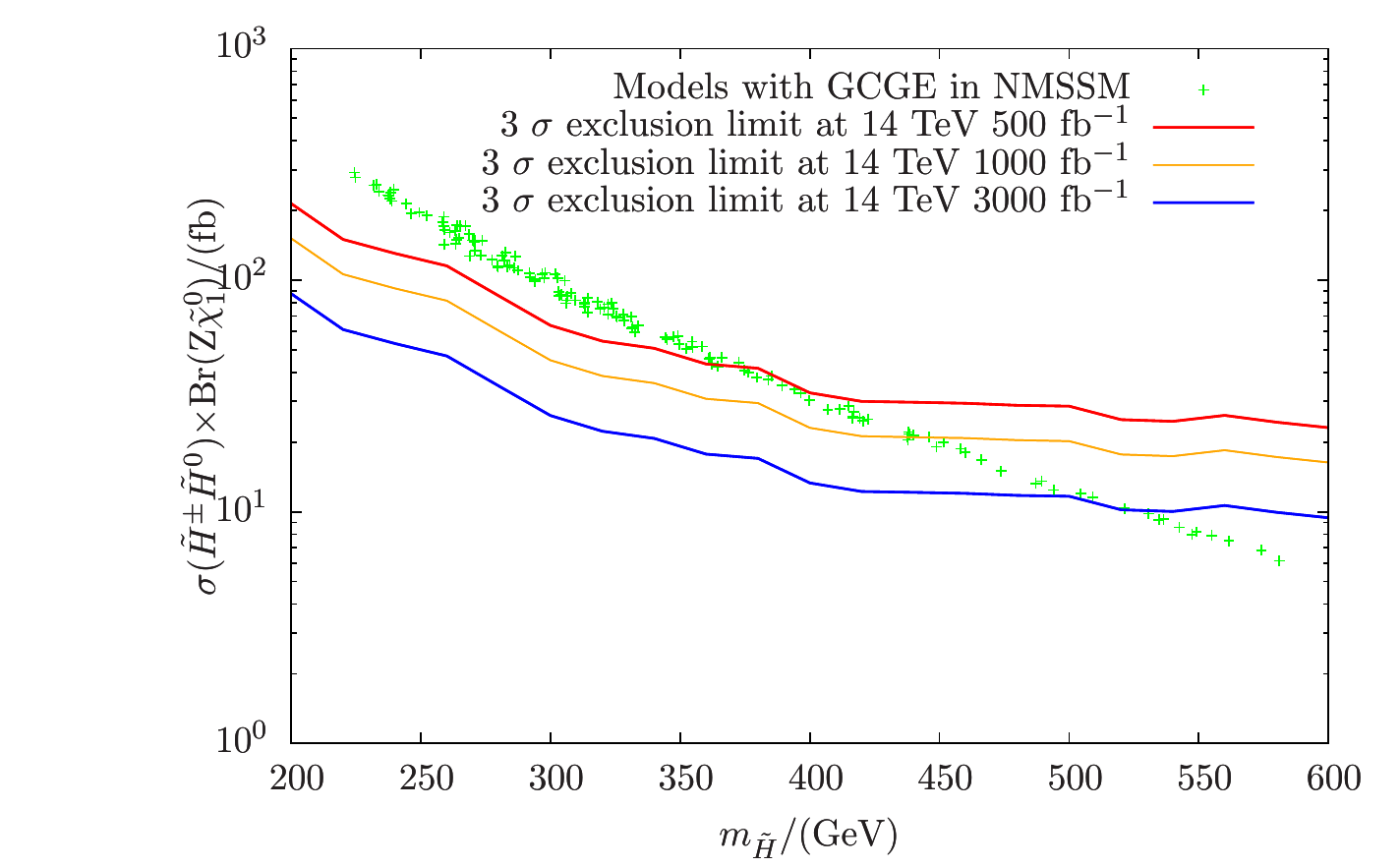}
\caption{\label{ewinowzh} Left: the production cross section of $WH$ final states and its corresponding exclusion limit 
at 14 TeV LHC with luminosity 3000 fb$^{-1}$. Right: the production cross sections of $WZ$ final states 
and their corresponding exclusion limits at 14 TeV LHC with luminosities 500 fb$^{-1}$, 1000 fb$^{-1}$ and 3000 fb$^{-1}$, respectively. }
\end{figure}

After obtaining the cut efficiency for the signal events, we can estimate the exclusion bound according to
\begin{align}
\sigma \sim \frac{S}{\sqrt{B}} = \frac{\sigma_s \epsilon_s}{ \sqrt{\sigma_b \epsilon_b}} \times{\sqrt{\mathcal{L}}}~,~
\end{align}
where $\sigma_b \epsilon_b$ is given in Table~\ref{whbg} and $\mathcal{L}$ is the luminosity. The corresponding 3-$\sigma$ exclusion 
limit for the $WH$ final state is shown as red curve in the left panel of Fig.~\ref{ewinowzh}.

The production cross section of the signal process for each of our NMSSM models is calculated by 
passing the SLHA~\cite{Allanach:2008qq} output that is generated from NMSSMtools into Madgraph5. 
We assume a K-factor of 1.2 for all those models.

Taking the resulting cross sections of the signal processes, we project our models to the exclusion plane 
that we have obtained before. The results are given in Fig.~\ref{ewinowzh} from which we conclude that 
only a small amount of our models can be detected at 14 TeV LHC with luminosity 3000 fb$^{-1}$. 
The relatively weak exclusion limit is mainly because of the low  cut efficiency of the signal events. We find the reduction of efficiency is dominated by the requirement of ``exactly two b-jets and one lepton in the final state", which reduces the number of signal events by about 2 order of magnitude. This can be roughly estimated by $\text{Br}(W\to l \nu) \times \text{Br}(h \to b \bar{b}) \times \epsilon_{\text{b tag}}^2 \sim 5 \times 10^{-2}$. In a realistic detector, the efficiency turns out to be further suppressed.
Furthermore, relatively strong cuts are imposed on missing transverse energy ($E^{\text{miss}}_T > 245$ GeV), contransverse mass ($M_{\text{CT}}> 160$ GeV) and transverse mass ($M_T>130$ GeV) in order to suppress the $Z$+jets/diboson, $t \bar{t}$ and $W$ backgrounds respectively. As a result, each of these cuts removes more than half of the events. 

\subsubsection{$WZ$ Channel}

The searches for both 3-lepton final states~\cite{Aad:2014nua,Baer:2012wg} and 2-lepton + jets~\cite{Aad:2014vma} give very strong bound 
on these decay modes of chargino and neutralio pair productions. The current bound is actually given by the combination 
of these two searches.   Even though the tri-lepton search is more sensitive to lighter chargino and neutralino region, 
the 2-lepton + jets is slightly better at higher mass region.
Thus, in this subsection, we will re-produce the SR-$Z$jets analysis in Ref.~\cite{Aad:2014vma} and extrapolate it to 14 TeV LHC with a few optimizations.
The cuts chosen by ATLAS Collaboration are listed as follows:
\begin{itemize}
\item Exact two opposite sign same flavor (OSSF) leptons are required. Also, these two leptons should have 
$p_T(l_1)> 35$ GeV and $p_T(l_2)> 20$ GeV, respectively;
\item Because the OSSF lepton pair is from a moderately boosted $Z$ boson decay, 
two additional cuts on leptons 
are imposed: $81.2~\text{GeV} < m_{ll} <101.2$~GeV, and $p_T(ll) > 80$ GeV. In addition, 
the angular separation between two leptons must satisfy $0.3< \Delta R(ll) < 1.5$;
\item There should be no $b$-jet, $\tau$-jet and forward jet in the event;
\item The two highest-$p_T$ central jets must have $p_T> 45$ GeV. They are expected from 
$W$ boson decay and then satisfy the invariant mass range $50~\text{GeV} < m_{jj} < 100$~GeV; and
\item A cut on $E^{\text{miss,rel}}_T$ is applied: $E^{\text{miss,rel}}_T > 80$ GeV, where
\begin{align}
E^{\text{miss,rel}}_T = \left\{
\begin{array}{l l}
E^{\text{miss}}_T & \text{, if } \Delta \phi_{l,j} \geq \pi/2\\
E^{\text{miss}}_T \times \Delta \phi_{l,j} & \text{, if } \Delta \phi_{l,j} < \pi/2
\end{array}~,~
\right.
\end{align}
where $\Delta \phi$ is the azimuthal angle between the direction of $\textbf{p}^{\text{miss}}_T$ and that of 
the nearest lepton or jet.
\end{itemize}

The dominant backgrounds are the leptonic decay of $W^\pm W^\mp$, the leptonic decay of $W^\pm Z$ 
where the lepton from $W$ decay is not reconstructed, and 
$ZZ$ with one of the $Z$ bosons decaying into charged leptons while the other decays into neutrinos.
All those backgrounds are generated with up to two additional jets by Madgraph5 using
a procedure similar to that described above.
In this case, the cross section before and after the cuts are given in Table~\ref{wzbg}.
 \begin{table}[htb]
 \small
 \begin{center}
  \begin{tabular}{c|c|c}
   &   $\sigma$(14 TeV)/pb   & SR/fb \\ \hline
   $W^\pm W^\mp$+jets & 124.3 & 0.029   \\
   $W^+ Z$+jets & 31.5 & 0.028 \\
   $W^- Z$+jets & 20.32 & 0.012  \\
   $ZZ$+jets & 17.72 & 0.13 \\
  \end{tabular}
  \caption{\label{wzbg} The background cross sections before and after the cuts for $WZ$ final state.}
 \end{center}
\end{table}

The signal process for the $WZ$ final state is generated similarly to what was done for the $WH$ process.
The expected 3-$\sigma$ exclusion limits for 14 TeV LHC with luminosities 500 fb$^{-1}$, 1000 fb$^{-1}$, 
and 3000 fb$^{-1}$ are given in the right panel of Fig.~\ref{ewinowzh}.
From the figure, we find the sensitivity for the $WZ$ channel is much better than that for the $WH$ channel, especially 
in the low mass region of the Higgsino. At 14 TeV with 500 fb$^{-1}$ from the LHC, the NMSSM models with Higgsino
lighter than $\sim 370$ GeV may be excluded.

\section{Conclusion}
\label{conc}

We have discussed  possible scenarios in the NMSSM with a discrete $Z_3$ symmetry that
may be able to explain the Galactic Centre Excess (GCE).
It turns out that the $s$-channel $A_1$ resonant annihilation into $b\bar{b}$, is 
the most promising scenario for realization of the the gamma-ray excess. In this scenario, even though the
GCE is produced by the $A_1$ resonant annihilation, 
the DM annihilation at freeze-out is mainly contributed to by the $s$-channel $Z$ boson exchange process. 
In order to get the correct DM relic density,  the coupling between the $Z$ boson and DM is enhanced by a 
relatively large Higgsino component in the DM and large tan$\beta$. We also briefly comment on the 
hidden sector dark matter scenario, in which DM is annihilating into a singlet Higgs pair. The subsequent decays of 
these Higgs bosons into $b$ quarks are able to produce the observed GCE signal. 
However,  in the NMSSM each of the DM mass, the DM coupling with Higgs bosons, and the Higgsino mass
are highly correlated. We need  a relatively large singlino dominant DM mass of $\sim 76$ GeV and a
larger superpotential coupling $\kappa$ to successfully reproduce
the GCE signal. As a result, the DM annihilation cross section is bounded from above by DM direct detection. 
So, the GCE can only be approximately fitted in hidden sector DM scenario.

The $s$-channel $A_1$ resonant process is the most promising channel to explain the GCE in the NMSSM. 
We have studied its LHC phenomenology for the parameter
space consistent with an explanation of the GCE. 
For this purpose we have considered the discovery potential of four promising signatures in detail. Firstly, 
we considered DM pair production which recoils
against a hard initial state radiation (ISR) jet. Because the energy spectrum of the ISR jet is 
suppressed by the small DM mass, the production rate 
of the required signal event is two orders of magnitude smaller than currently available sensitivity. 
The situation will not become significantly better for the next higher-energy phase of the LHC. 
Secondly, we considered the pair production of neutral Higgsinos that subsequently 
decay into a Higgs pair. This channel, despite its interesting features, is two orders of magnitude smaller
 than current sensitivity as well. 
Thirdly, we considered two traditional channels which consist of chargino and neutralino pair production
with subsequent decays into $WH$+DMs and $WZ$+DMs, respectively. By extrapolating the
two most sensitive analyses of the ATLAS Collaboration, we found that for the LHC with 14 TeV energy:
(a) for the $WH$ channel part of the model space for the first scenario is discoverable with an integrated luminosity 
of 3000 fb$^{-1}$; and (b) for the $WZ$ channel most of the model space is discoverable and only requires 500 fb$^{-1}$
of integrated luminosity.\\ \\

\section*{Acknowledgements}

We would like to thank Yandong Liu for collaboration at the early stage of the work. 
This research was supported in part by
 the Australian Research Council through Grant No. CE110001004 (CoEPP) and by 
the University of Adelaide (JL and AGW), as well as by the Natural Science
Foundation of China under grant numbers 10821504, 11075194, 11135003, 11275246, and 11475238,
and by the National Basic Research Program of China (973 Program) 
under grant number 2010CB833000 (JG and TL).

\appendices
\section{Neutralino and Higgs Mass Matrices}
\label{NHmass}
We use the convention in Ref.~\cite{Ellwanger:2009dp}
\begin{align}
\text{tan} \beta &= \frac{v_u}{v_d}, \\
\mu_{\text{eff}} &= \lambda v_s, \\
M_A^2 &= \frac{2 \lambda v_s}{\text{sin} 2 \beta} (A_\lambda + \kappa v_s),\\
v = \sqrt{v_u^2+v_d^2} & = 174 ~\text{GeV}~,
\end{align}
where $v_d$, $v_u$, and $s$ are the VEVs of $H_d$, $H_u$, and $S$, respectively.
The symmetric neutralino mass matrix in the ($\tilde{B}, \tilde{W}, \tilde{H_d}, \tilde{H_u}, \tilde{S}$) basis 
can be written as follows
\begin{align}
\mathcal{M}_0 =
\begin{pmatrix}
M_1 & 0 & - \frac{g_1 v_d}{\sqrt{2}} & \frac{g_1 v_u}{\sqrt{2}} & 0 \\
       & M_2 & \frac{g_2 v_d}{\sqrt{2}} & -\frac{g_2 v_u}{\sqrt{2}} & 0 \\
       & & 0 & - \mu_{\text{eff}} & - \lambda v_u \\
       & & & 0 & -\lambda v_d \\
       & & & & 2 \kappa s \\
\end{pmatrix}.
\end{align}
The lightest neutralino is the mixing of gauge eigenstates 
\begin{align}
\tilde{\chi}^0_1 = N_{11} \tilde{B} + N_{12} \tilde{W} + N_{13} \tilde{H_u} + N_{14} \tilde{H_d} + N_{15} \tilde{S},
\end{align}
and its mass approximately is $m_{\tilde{\chi}^0_1} \sim 2 \kappa s$ if it is singlino-like.
In the $2 \kappa/ \lambda \ll 1$ limit, the Higgsino components of the $\tilde{\chi}^0_1$ can be estimated as
below~\cite{Cheung:2014lqa}
\begin{align}
\frac{N_{13}}{N_{15}} & = \frac{\lambda v}{\mu^2 - m^2_{\tilde{\chi}^0_1}} c_\beta (t_\beta ~m_{\tilde{\chi}^0_1} - \mu ), \label{hdcomp}\\
\frac{N_{14}}{N_{15}} & = \frac{- \lambda v}{\mu^2 - m^2_{\tilde{\chi}^0_1} } s_\beta (\mu - \frac{m_{\tilde{\chi}^0_1}}{t_\beta})~, \label{hucomp}
\end{align}
where we have used $c_\beta$, $s_\beta$, and $t_\beta$ for $\text{cos}\beta$, $\text{sin}\beta$, and $\text{tan}\beta$, respectively.

The $3 \times 3$ symmetric CP-even Higgs boson mass matrix in the $(S_1, S_2, S_3)$ basis~\cite{Miller:2003ay} is
\begin{align}
\mathcal{M}^2_S =
\begin{pmatrix}
M^2_A + s^2_{2 \beta} (m^2_Z - \lambda^2 v^2)  & ~~~ s_{2\beta} c_{2 \beta} (m^2_Z - \lambda^2 v^2) & ~~~  - \lambda v c_{2 \beta}(A_\lambda + \frac{2 \kappa \mu}{\lambda}) \\
  & c^2_{2\beta} m^2_Z + \lambda^2 v^2 s^2_{2\beta} &~~~ 2\lambda v (\mu - s_\beta c_\beta(A_\lambda + \frac{2 \kappa \mu}{\lambda})) \\
  & &  \frac{s_{2 \beta}}{2} \frac{\lambda^2 v^2}{\mu} A_\lambda + \frac{\kappa \mu}{\lambda} (A_\kappa + 4 \frac{\kappa \mu}{\lambda})\\
\end{pmatrix}.
\end{align}
In the case of very light singlet, the singlet component of the SM Higgs $H_{\text{SM}}$ can be suppressed by requiring
\begin{align}
(\mathcal{M}^2_S)_{23} \simeq 0.
\label{m23e0}
\end{align}
The Eq.~(\ref{m23e0}) can be used to determine the approximation value of $A_\lambda$ as follows
\begin{align}
A_\lambda \simeq \frac{2 \mu}{s_{2 \beta}} - \frac{2 \kappa \mu}{\lambda}.
\label{alambda}
\end{align}

Finally, we give  the simplest $2 \times 2$ symmetric CP-odd Higgs mass matrix in the $(P_1, P_2)$ basis
\begin{align}
\mathcal{M}^2_P =
\begin{pmatrix}
M^2_A & \lambda v (A_\lambda - 2 \frac{\kappa \mu}{\lambda})  \\
 & -\frac{3 \kappa \mu A_\kappa}{\lambda} + \frac{\lambda^2 v^2}{2 \mu} s_{2 \beta}(A_\lambda + \frac{4 \kappa \mu}{\lambda})
\end{pmatrix}~.~
\label{pmass}
\end{align}

\bibliography{NMSSM_GCE}
\bibliographystyle{utphys}

\end{document}